
\input jytex
\typesize=12pt
\sectionnumstyle{arabic}
\footnoteskip=2pt
\def\NPB#1#2#3{{\sl Nucl. Phys.} {\bf B#1} (#2) #3}
\def\PRL#1#2#3{{\sl Phys. Rev. Lett.} {\bf#1} (#2) #3}
\def\Phy#1#2#3{{\sl Physica} {\bf #1} (#2) #3}
\def\JPA#1#2#3{{\sl J. Physics} {\bf A#1} (#2) #3}
\def\JETP#1#2#3{{\sl Sov. Phys. JETP} {\bf #1} (#2) #3}
\def\JSM#1#2#3{{\sl J. Soviet Math.} {\bf #1} (#2) #3}
\catcode`\@=11

\catcode`\^^?=13 \def^^?{\relax}

\def\Odoubled#1{{\setbox0=\hbox{\rm#1}%
     \dimen@=\ht0 \dimen@ii=.04em \advance\dimen@ by-\dimen@ii
     \rlap{\kern.26em \vrule height\dimen@ depth-\dimen@ii width.075em}\box0}}
\def\Complex{\Odoubled C}
\def\upref#1/{\markup{[\putref{#1}]}}
\def\tqd{{\tilde Q}^\dagger}
\def\cd{c^\dagger}

\def\half{{1\over 2}}
\def\up{\uparrow}
\def\down{\downarrow}
\def\vac{|0\rangle}
\def\loc{|\down\up\rangle}
\def\vacn{|0\rangle^{(1)}}
\def\vacnn{|0\rangle^{(2)}}
\def\l#1{\lambda_{#1}}
\def\tl#1{{\tilde\lambda}_{#1}}
\def\tonel#1{{\tilde\la}^{(1)}_{#1}}
\def\ttwol#1{{\tilde\la}^{(2)}_{#1}}
\def\la{\lambda}
\def\1l#1{\lambda^{(1)}_{#1}}
\def\2l#1{\lambda^{(2)}_{#1}}
\def\a#1#2{a({#1}-{#2})}
\def\b#1#2{b({#1}-{#2})}
\def\ta#1{{{#1}-{i\over 2}\over {#1}+{i\over 2}}}
\def\tb#1{{{#1}+{i\over 2}\over {#1}-{i\over 2}}}
\def\tc#1{{{#1}+{i}\over {#1}-{i}}}
\def\P{\Pi}
\def\Pg{\Pi}
\def\pg{\Pi^{(1)}}
\def\ppg{\Pi^{(2)}}
\def\pbbf{\Pi_{BBF}}
\def\pffb{\Pi_{FFB}}
\def\LE#1#2#3#4#5#6{{L^{(1)}_{#1}(#2 -\l{#1})}^{#3#4}_{#5#6}}
\def\LEn#1#2#3#4#5#6{{L^{(2)}_{#1}(#2 -\1l{#1})}^{#3#4}_{#5#6}}
\def\Le1#1#2{L^{(1)}_{#1}(#2 -\l{#1})}
\def\Len#1#2{L^{(2)}_{#1}(#2 -\1l{#1})}
\def\Let#1#2{L^{(2)}_{#1}(#2 -\1l{#1})}
\def\ra#1#2#3#4#5#6{{r(#1-#2)^{#3#4}_{#5#6}\over \a#1#2 }}
\def\ran#1#2#3#4#5#6{{r^{(1)}(#1-#2)^{#3#4}_{#5#6}\over \a#1#2 }}
\def\r#1#2#3#4#5#6{r(#1-#2)^{#3#4}_{#5#6}\ }
\def\rffb#1#2#3#4#5#6{r_{\sscr FFB}(#1-#2)^{#3#4}_{#5#6}\ }
\def\rn#1#2#3#4#5#6{r^{(1)}(#1-#2)^{#3#4}_{#5#6}\ }
\def\rffb#1#2#3#4#5#6{r_{FFB}(#1-#2)^{#3#4}_{#5#6}\ }

\def\R#1#2#3#4#5#6{R(#1-#2)^{#3#4}_{#5#6}\ }
\def\sr#1#2#3#4#5{r(#1)^{#2#3}_{#4#5}\ }
\def\srffb#1#2#3#4#5{r_{FFB}(#1)^{#2#3}_{#4#5}\ }
\def\srbbf#1#2#3#4#5{r_{BBF}(#1)^{#2#3}_{#4#5}\ }
\def\sR#1#2#3#4#5{R(#1)^{#2#3}_{#4#5}\ }
\def\ga{\alpha}
\def\gb{\beta}
\def\gc{\gamma}
\def\gd{\delta}
\def\ge{\epsilon}
\def\gs{\sigma}

\def\vmn{V^{(m|n)}}
\def\v12{V^{(2|2)}}
\def\osl{\bar\otimes}
\def\scr{\scriptstyle}

\def\da{\downarrow}
\def\pl#1{(\putlab{#1})}
\equation{config}{\vac \ , \qquad
|\uparrow\rangle _i = c^\dagger_{i,1} \vac \ , \qquad
|\downarrow\rangle _i = c^\dagger_{i,-1} \vac,\qquad
|\down\up\rangle _i = c^\dagger_{i,-1} c^\dagger_{i,1}\vac
\quad .}

\equation{hnaught}{\eqalign{H^0_{j,j+1} &=
\cd_{{j+1},{1}} c_{j,{1}}(1-n_{j,{-1}}-n_{{j+1},{-1}})
+ \cd_{j,{1}} c_{{j+1},{1}}(1-n_{j,{-1}}-n_{{j+1},{-1}})
\cr
& + \cd_{{j+1},{-1}} c_{j,{-1}}(1-n_{j,{1}}-n_{{j+1},{1}})
+ \cd_{j,{-1}} c_{{j+1},{-1}}(1-n_{j,{1}}-n_{{j+1},{1}})
\cr
& + \half (n_j - 1)(n_{j+1} - 1)
+ \cd_{j,{1}} \cd_{j,{-1}} c_{{j+1},{-1}} c_{{j+1},{1}}
+ c_{j,{-1}} c_{j,{1}} \cd_{{j+1},{1}} \cd_{{j+1},{-1}}
\cr
&  - \half (n_{j,{1}}-n_{j,{-1}})(n_{{j+1},{1}}-n_{{j+1},{-1}})
- \cd_{j,{-1}} c_{j,{1}} \cd_{{j+1},{1}} c_{{j+1},{-1}}
- \cd_{j,{1}} c_{j,{-1}} \cd_{{j+1},{-1}} c_{{j+1},{1}}
\cr
&  + (n_{j,{1}}-\half)(n_{j,{-1}}-\half)
+ (n_{{j+1},{1}}-\half)(n_{{j+1},{-1}}-\half) \ .\cr}}

\equation{h}{H=H^0+U\sum_{j=1}^L (n_{j,1}-{1\over 2})(n_{j,-1}-{1\over
2}) -\mu\sum_{j=1}^L n_j + h\sum_{j=1}^L (n_{j,1}-n_{j,-1})}

\equation{spin}{
S_j = c^\dagger_{j,1} c_{j,-1}\ , \qquad
S^\dagger_j = c^\dagger_{j,-1} c_{j,1}\ ,\qquad
S^z_j = {1\over 2} (n_{j,1}-n_{j,-1}) \quad ,}

\equation{eta}{
\eta_j = c_{j,{1}} c_{j,{-1}}\ , \qquad
\eta^\dagger_j = c^\dagger_{j,{-1}} c^\dagger_{j,{1}}\ , \qquad
\eta^z_j = - \half n_j + \half\quad .}

\equation{perm}{\eqalign{
\Pg^{j,j+1}\vac _j\times \vac _{j+1} &=
\vac _{j}\times \vac _{j+1}\cr
\Pg^{j,j+1}\loc _j\times \loc _{j+1} &=
\loc _{j}\times \loc _{j+1}\cr
\Pg^{j,j+1}\vac _j\times |\sigma\rangle_{j+1} &=
|\sigma\rangle_{j}\times \vac _{j+1}\cr
\Pg^{j,j+1}|\tau\rangle_{j}\times |\sigma\rangle_{j+1} &=
- |\sigma\rangle_{j}\times |\tau\rangle_{j+1}
\quad ,\qquad \sigma ,\tau =\uparrow
,\downarrow\cr
& {\sl etc.}\cr}}

\equation{X}{X=\sum_{j=1}^L X_j\ ,\quad X_j = (n_{j,1} -{1\over
2})(n_{j,-1} -{1\over 2})\quad . }

\equation{ferm}{Q_{j,\gs} = (1-n_{j,-\gs}) c_{j,\gs}\ ,\quad
{\tilde Q}_{j,\gs} = n_{j,-\gs} c_{j,\gs}\ ,\ \gs =\pm 1}

\equation{hintro}{H_{(3)} = i\sum_{k=1}^L \lbrack H^0_{k,k+1},
H^0_{k-1,k}\rbrack\quad ,}

\equation{hpi}{H^0 = -\sum_{j=1}^L \Pg^{j,j+1}\quad ,}

\equation{linops}{M=\left(\matrix{A&B\cr C&D\cr}\right)\ ,\quad
\ge\left(\matrix{A&0\cr 0&D\cr}\right)=0\ ,\quad
\ge\left(\matrix{0&B\cr C&0\cr}\right)=1\quad .}

\equation{str}{str(M) = \sum_{a=1}^{m+n} (-1)^{\ge_a} M_{aa}\quad .}

\equation{tensorprod}{v\otimes w = (e_av_a)\otimes (e_bw_b) =
(e_a\otimes e_b)v_aw_b (-1)^{\ge_{v_a}\ge_b}\quad .}

\equation{VV}{\left(R(\l{})\ e_j\otimes e_k\right)^{a_1}_{a_2} =
e_j^{b_1}\otimes e_k^{b_2}\ R(\l{})^{b_1a_1}_{b_2a_2}
\quad .}

\equation{linop}{\left(F\otimes G\right)(v\otimes w) = F(v)\otimes
G(w)\quad ,}

\equation{linopmat}{\left(F\otimes G\right)^{ab}_{cd} = F_{ab}G_{cd}\
(-1)^{\ge_c(\ge_a+\ge_b)} \quad .}

\equation{pigr}{\eqalign{
I (v\otimes w) &= (v\otimes w),\quad
(I)^{a_1 b_1}_{a_2 b_2} = \delta_{a_1b_1}\delta_{a_2b_2}\cr
\Pg (v\otimes w) &= (w\otimes v),\quad
(\Pg)^{a_1 b_1}_{a_2 b_2} = \delta_{a_1b_2}
\delta_{a_2 b_1} (-1)^{\epsilon_{b_1}\epsilon_{b_2}}\quad .\cr}}

\equation{YBE}{\R{\l{}}{\mu}{a_2}{c_2}{a_3}{c_3}
\sR{\l{}}{a_1}{b_1}{c_2}{d_2}\sR{\mu}{d_2}{b_2}{c_3}{b_3} =
\sR{\mu}{a_1}{c_1}{a_2}{c_2}\sR{\l{}}{c_2}{d_2}{a_3}{b_3}
\R{\l{}}{\mu}{c_1}{b_1}{d_2}{b_2}\quad .}

\equation{YBE1}{\left(I\otimes R(\la -\mu)\right)
\left(R(\la )\otimes I\right)
\left(I\otimes R(\mu )\right) =
\left(R(\mu )\otimes I\right)
\left(I\otimes R(\la )\right)
\left(R(\la -\mu )\otimes I\right)
\quad .}

\equation{Rgen}{\eqalign{&R({\l{}}) = b(\l{})I + a(\l{}) \Pg\cr
&a(\l{}) = {\l{}\over \l{} +i}\ ,\qquad b(\l{}) = {i\over \l{}
+i}\quad .}}

\equation{int}{\eqalign{&\R{\l{}}{\mu}{f_1}{c_1}{e_2}{c_2}\
\left(\Pg \ R(\l{})\right)^{c_1b_1}_{f_3 c_3}\
\left(\Pg \ R(\mu)\right)^{c_2 b_2}_{c_3 b_3}\
(-1)^{\ge_{c_2}(\ge_{c_1}+\ge_{b_1})} = \cr
&\left(\Pg \ R(\mu)\right)^{f_1 c_1}_{f_3 c_3}\
\left(\Pg \ R(\l{})\right)^{e_2c_2}_{c_3 b_3}\
\R{\l{}}{\mu}{c_1}{b_1}{c_2}{b_2}\
(-1)^{\ge_{e_2}(\ge_{f_1}+\ge_{c_1})}
\qquad .\cr}}

\equation{YBE2}{R_{12}(\l{}-\mu)\left(\left\lbrack\P_{13}
R_{13}(\l{})\right\rbrack\otimes \left\lbrack\P_{23}
R_{23}(\mu)\right\rbrack\right) =
\left(\left\lbrack\P_{13} R_{13}(\mu)\right\rbrack\otimes
\left\lbrack\P_{23} R_{23}(\l{})\right\rbrack\right)
 R_{12}(\l{}-\mu)\quad ,}

\equation{LR}{L_n(\l{})^{ab}_{\alpha\beta}=\Pg^{ac}_{\ga\gc}
\sR{\l{}}{c}{b}{\gc}{\gb} = \left( b(\la )\Pg + a(\la )
I\right)^{ab}_{\ga\gb} \qquad .}

\equation{int1}{\eqalign{&\R{\l{}}{\mu}{a_1}{c_1}{a_2}{c_2}\
L_n(\l{})^{c_1b_1}_{\ga _n \gc_n}\
L_n(\mu)^{c_2 b_2}_{\gamma _n\beta _n}\
(-1)^{\ge_{c_2}(\ge_{c_1}+\ge_{b_1})}\ = \cr
&L_n(\mu)^{a_1 c_1}_{\ga _n \gc _n}\
L_n(\l{})^{a_2c_2}_{\gamma _n \beta _n}\
(-1)^{\ge_{a_2}(\ge_{a_1}+\ge_{c_1})}\
\R{\l{}}{\mu}{c_1}{b_1}{c_2}{b_2}\qquad .\cr}}

\equation{int2}{R(\l{}-\mu)\ \left(L_n(\l{})\otimes
L_n(\mu)\right) = \left(L_n(\mu)\otimes
L_n(\l{})\right)\ R(\l{}-\mu)\quad .}

\equation{mon}{\eqalign{T_L(\l{}) &= L_L(\l{})L_{L-1}(\l{})\ldots
L_1(\l{})\cr
\left((T_L(\l{}))^{ab}\right)_{{\scriptstyle \ga_1\ldots\ga_L}\atop
{\scriptstyle \gb_1\ldots\gb_L}} &= L_L(\l{})^{ac_L}_{\ga_L\gb_L}
L_{L-1}(\l{})^{c_Lc_{L-1}}_{\ga_{L-1}\gb_{L-1}}\ldots
L_{1}(\l{})^{c_2b}_{\ga_{1}\gb_{1}}\quad\times\cr &\quad\times
(-1)^{\sum_{j=2}^L(\ge_{\ga_j}+\ge_{\gb_j})\sum_{i=1}^{j-1}
\ge_{\ga_i}}\quad .\cr}}

\equation{intT}{R(\l{}-\mu)\ \left(T_L(\l{})\otimes
T_L(\mu)\right) = \left(T_L(\mu)\otimes
T_L(\l{})\right)\ R(\l{}-\mu)\quad .}

\equation{tau}{\tau (\l{}) = str(T_L(\l{})) = \sum_{a=1}^{m+n}
(-1)^{\ge_a} (T_L(\la ))^{aa}\qquad .}

\equation{trid2}{H^0 = -i {\partial \log(\tau (\l{}))\over
\partial\l{}}\bigg |_{\l{}=0} - L\ =\ H_{(2)} -L\quad ,}

\equation{trid1}{H_{(2)} = -i\ {\partial \log(\tau (\l{}))\over
\partial\l{}}\bigg |_{\l{}=0} = -{\sum_{k=1}^L (\Pg^{k,k+1}-1)}.}

\equation{hcl}{\log\left(\tau(\l{})(\tau (0))^{-1}\right) =
\sum_{k=1}^\infty i{\l{}^k\over k!}
H_{(k+1)}\qquad .}

\equation{mtm}{P =-i\ \log\left(\tau (0)\right)\qquad .}


\equation{vacuum}{{\vac _n = \pmatrix{0\cr 0\cr 0\cr 1\cr}\qquad
,\qquad \vac = \otimes_{n=1}^L \vac _n}\quad .}

\equation{Lop}{L_n(\l{}) =
\left({\matrix{
a(\l{})-b(\l{})e_n^{11}&-b(\l{})e_n^{21}&b(\l{})e_n^{31}&b(\la)e_n^{41}\cr
-b(\l{})e_n^{12}&a(\l{})-b(\l{})e_n^{22}&b(\l{})e_n^{32}&b(\la)e_n^{42}\cr
b(\l{})e_n^{13}&b(\l{})e_n^{23}&a(\l{})+b(\la)e_n^{33}&b(\la)e_n^{43}\cr
b(\l{})e_n^{14}&b(\l{})e_n^{24}&b(\la)e_n^{34}&a(\la)+b(\la)e_n^{44}\cr
}}\right)\quad ,}

\equation{T}{{T_L(\l{}) = L_L(\l{})L_{L-1}(\l{})\ldots
L_1(\l{})=\left(\matrix{
A_{11}(\l{})&A_{12}(\l{})&A_{13}(\l{})&B_1(\l{})\cr
A_{21}(\l{})&A_{22}(\l{})&A_{23}(\l{})&B_2(\l{})\cr
A_{31}(\l{})&A_{32}(\l{})&A_{33}(\l{})&B_3(\l{})\cr
C_1(\l{})&C_2(\l{})&C_3(\l{})&D(\l{})\cr}\right)}\quad .}

\equation{tau2}{\tau (\mu) = str(T_L(\mu )) = -A_{11}(\mu )-A_{22}(\mu
)+ A_{33}(\mu)+D(\mu )\qquad .}

\equation{Lvac}{L_k(\l{})\vac _k = \left(\matrix{a(\l{})&0&0&0\cr
0&a(\l{})&0&0\cr 0&0&a(\la)&0\cr
b(\l{})e_k^{14}&b(\l{})e_k^{24}&b(\l{})e_k^{34}&1}\right)\vac _k\qquad
.}

\equation{Tvac}{T_L(\l{})\vac = \left(\matrix{(a(\l{}))^L&0&0&0\cr
0&(a(\l{}))^L&0&0\cr 0&0&(a(\l{}))^L&0\cr
C_1(\l{})&C_2(\l{})&C_3(\la)&1}\right)\vac\qquad .}

\equation{state1}{|\l{1},\ldots , \l{n}|F\rangle = C_{a_1}(\l{1})\
C_{a_2}(\l{2})\ldots C_{a_n}(\l{n})\ \vac\  F^{a_n\ldots
a_1}\qquad ,}

\equation{AC}{\eqalign{
A_{ab}(\mu)\ C_c(\l{}) &= (-1)^{\ge_a\ge_p}\
\ra{\mu}{\la}{d}{c}{p}{b}\ C_p(\l{})\ A_{ad}(\mu)\cr
&\qquad -(-1)^{\ge_a\ge_b}\ {\b{\mu}{\la}\over\a{\mu}{\la}}
C_{b}(\mu) A_{ac}(\la )\cr
D(\mu)\ C_c(\l{}) &=
{1\over \a{\la}{\mu}}\ C_c(\l{})\ D(\mu)\ -\ {\b{\la}{\mu}\over
\a{\la}{\mu}} C_{c}(\mu ) D(\la )\cr
C_{a_1}(\l{1})\ C_{a_2}(\l{2}) &=
\r{\l{1}}{\l{2}}{b_1}{a_2}{b_2}{a_1}\ C_{b_2}(\l{2})\
C_{b_1}(\l{1})\quad ,\cr}}

\equation{r}{\eqalign{\sr{\mu}{a}{b}{c}{d} &= b({\mu})\delta_{ab}
\delta_{cd} + a({\mu})\delta_{ad}\delta_{bc}(-1)^{\ge_a\ge_c}\cr
&= b({\mu}){I^{(1)}}^{ab}_{cd} + a({\mu}){\pg}^{ab}_{cd}\quad .\cr}}

\equation{ybe}{\r{\l{}}{\mu}{a_2}{c_2}{a_3}{c_3}
\sr{\l{}}{a_1}{b_1}{c_2}{d_2}\sr{\mu}{d_2}{b_2}{c_3}{b_3} =
\sr{\mu}{a_1}{c_1}{a_2}{c_2}\sr{\l{}}{c_2}{d_2}{a_3}{b_3}
\r{\l{}}{\mu}{c_1}{b_1}{d_2}{b_2}\qquad ,}

\equation{Dstate}{\eqalign{D(\mu )|\l{1},\ldots , \l{n}|F\rangle =
&\prod_{j=1}^n {1\over \a{\l{j}}{\mu}} |\l{1},\ldots , \l{n}|F\rangle\cr
&+\sum_{k=1}^n\left({\tilde\Lambda}_k\right)^{b_1\ldots b_n}_{a_1\ldots a_n}
C_{b_k}(\mu )\prod_{\scr j=1\atop\scr j\neq k}^n C_{b_j}(\l{j})\vac
F^{a_n\ldots a_1}\quad ,\cr}}

\equation{Astate}{\eqalign{
&(-A_{11}(\mu )-A_{22}(\mu )+A_{33}(\mu))|\l{1},\ldots ,
\l{n}|F\rangle =\cr
&\quad=\ (a(\mu ))^L\prod_{j=1}^n {1\over \a{\mu}{\l{j}}}
\prod_{l=1}^n C_{b_l}(\l{l})\vac\ \tau^{(1)}(\mu )^{b_1\ldots
b_n}_{a_1\ldots a_n}\ F^{a_n\ldots a_1} \cr
&\quad\qquad +\sum_{k=1}^n \left(\Lambda_k\right)^{b_1\ldots
b_n}_{a_1\ldots a_n} C_{b_k}(\mu )
\prod_{\scr j=1\atop\scr j\neq k}^n
C_{b_j}(\l{j})\vac F^{a_n\ldots a_1}\quad ,\cr}}

\equation{tau1}{\eqalign{\tau^{(1)}(\mu )^{b_1\ldots
b_n}_{a_1\ldots a_n} &= str( T_n^{(1)}(\mu )) \cr
&= str(\Le1{n}{\mu}\Le1{n-1}{\mu}\ldots\Le1{2}{\mu}\Le1{1}{\mu})\quad
,\cr
&{\hskip -70pt\rm and}\cr
L_k^{(1)}(\la) &= b(\la) \pg + a(\la) I^{(1)}\cr
&= \pg\ r(\la) =
\left({\matrix{
a(\l{})-b(\l{})e_n^{11}&-b(\l{})e_n^{21}&b(\l{})e_n^{31}\cr
-b(\l{})e_n^{12}&a(\l{})-b(\l{})e_n^{22}&b(\l{})e_n^{32}\cr
b(\l{})e_n^{13}&b(\l{})e_n^{23}&a(\l{})+b(\la
)e_n^{33}\cr}}\right)\quad .\cr}}

\equation{ew}{\tau (\mu)|\l{1},\ldots , \l{n}|F\rangle = \nu
(\mu)\ |\l{1},\ldots , \l{n}|F\rangle }

\equation{unw}{\left((\Lambda_k)^{b_1\ldots b_n}_{a_1\ldots a_n} +
(\tilde\Lambda_k)^{b_1\ldots b_n}_{a_1\ldots a_n}\right)
F^{a_n\ldots a_1} = 0\quad .}

\equation{PBC}{(a(\l{k}))^{-L} \prod_{\scr
l=1\atop\scr l\neq k}^n
{\a{\l{k}}{\l{l}}\over\a{\l{l}}{\l{k}}}\ F^{b_n\ldots b_1} =
\tau^{(1)}(\l{k})^{b_1\ldots b_n}_{a_1\ldots a_n}\ F^{a_n\ldots
a_1}\quad ,\quad k=1,\ldots ,n\quad .}

\equation{intnest}{r(\l{}-\mu)\left(T_n^{(1)}(\l{})\otimes
T_n^{(1)}(\mu)\right)=\left(T_n^{(1)}(\mu)\otimes
T_n^{(1)}(\l{})\right) r(\l{}-\mu)\quad .}

\equation{T1}{{T^{(1)}_n(\mu) =
\left(\matrix{A^{(1)}_{11}(\mu)&A^{(1)}_{12}(\mu)&B^{(1)}_1(\mu)\cr
A^{(1)}_{21}(\mu)&A^{(1)}_{22}(\mu)&B^{(1)}_2(\mu)\cr
C^{(1)}_1(\mu)&C^{(1)}_2(\mu)&D^{(1)}(\mu)\cr}\right),\
\tau^{(1)}(\mu) = -A^{(1)}_{11}(\mu )-A^{(1)}_{22}(\mu ) + D^{(1)}(\mu
)\ ,}}

\equation{intnest2}{\eqalign{A^{(1)}_{ab}(\mu)\ C^{(1)}_c(\l{}) =&
(-1)^{\ge_a\ge_p}\ \ran{\mu}{\la}{d}{c}{p}{b}\ C^{(1)}_p(\l{})\
A^{(1)}_{ad}(\mu)\cr
&\qquad - (-1)^{\ge_a\ge_b}\ {\b{\mu}{\la}\over \a{\mu}{\la}}
C^{(1)}_{b}(\mu) A^{(1)}_{ac}(\la )\cr D^{(1)}(\mu)\ C^{(1)}_c(\l{})
&=  {1\over \a{\la}{\mu}}\ C^{(1)}_c(\l{})\ D^{(1)}(\mu)\ -\
{\b{\la}{\mu}\over \a{\la}{\mu}} C^{(1)}_{c}(\mu ) D^{(1)}(\la )\cr
C^{(1)}_{a_1}(\l{1})\ C^{(1)}_{a_2}(\l{2}) &=
r^{(1)}(\l{1}-\l{2})^{b_1a_2}_{b_2a_1}\ C^{(1)}_{b_2}(\l{2})\
C^{(1)}_{b_1}(\l{1})\quad .\cr}}

\equation{vacnest}{{\vac ^{(1)}_k = \pmatrix{0\cr 0\cr 1\cr}\qquad
,\qquad \vac ^{(1)} = \otimes_{k=1}^n \vac ^{(1)}_k}\quad .}

\equation{tvac}{\eqalign{A^{(1)}_{11}(\mu)\vac ^{(1)} &=
A^{(1)}_{22}(\mu)\vac ^{(1)} = \prod_{j=1}^n
\a{\mu}{\l{j}}\vacn\cr
D^{(1)}(\mu)\vac ^{(1)} &= \vacn\quad .\cr}}

\equation{state2}{|\1l{1},\ldots , \1l{n_1}|G\rangle =
C^{(1)}_{b_1}(\1l{1})\  C^{(1)}_{b_2}(\1l{2})\ldots
C^{(1)}_{b_{n_1}}(\1l{n_1})\ \vacn\  G^{b_{n_1}\ldots b_1}\qquad .}

\equation{Dstate1}{\eqalign{D^{(1)}(\mu )|\1l{1},\ldots ,
\1l{n_1}|G\rangle =
&\prod_{j=1}^{n_1} {1\over \a{\1l{j}}{\mu}} |\1l{1},\ldots ,
\1l{n_1}|G\rangle\cr
&+\sum_{k=1}^{n_1}\left({\tilde\Lambda}^{(1)}_k\right)^{b_1\ldots
b_{n_1}}_{a_1\ldots a_{n_1}}
C^{(1)}_{b_k}(\mu )\prod_{\scr j=1\atop\scr j\neq k}^{n_1}
C^{(1)}_{b_j}(\1l{j})\vacn G^{a_{n_1}\ldots a_1}\quad ,\cr}}

\equation{Astate1}{\eqalign{
&(-A^{(1)}_{11}(\mu )-A^{(1)}_{22}(\mu ))|\1l{1},\ldots ,
\1l{n_1}|G\rangle =\cr
&\qquad =\ \prod_{l=1}^n a(\mu -\l{l})\
\prod_{j=1}^{n_1} {1\over \a{\mu}{\1l{j}}}
\prod_{l=1}^{n_1} C^{(1)}_{b_l}(\1l{l})\vacn\ \tau^{(2)}(\mu )^{b_1\ldots
b_{n_1}}_{a_1\ldots a_{n_1}}\ G^{a_{n_1}\ldots a_1} \cr
&\qquad\qquad +\sum_{k=1}^{n_1} \left(\Lambda^{(1)}_k\right)^{b_1\ldots
b_{n_1}}_{a_1\ldots a_{n_1}} C^{(1)}_{b_k}(\mu )
\prod_{\scr j=1\atop\scr j\neq k}^{n_1}
C^{(1)}_{b_j}(\1l{j})\vacn G^{a_{n_1}\ldots a_1}\quad .\cr}}

\equation{tau3}{\eqalign{\tau^{(2)}(\mu )^{b_1\ldots
b_{n_1}}_{a_1\ldots a_{n_1}} &= str( T_{n_1}^{(2)}(\mu )) \cr
&= str(\Let{{n_1}}{\mu}\Let{{n_1}-1}{\mu}\ldots\Let{2}{\mu}\Let{1}{\mu})\quad
,\cr
&{\hskip -70pt\rm where}\cr
L_k^{(2)}(\la) &= b(\la) \ppg + a(\la) I^{(2)}\cr
&= \ppg\ r^{(1)}(\la) =
\left({\matrix{
a(\l{})-b(\l{})e_{k}^{11}&-b(\l{})e_{k}^{21}\cr
-b(\l{})e_{k}^{12}&a(\l{})-b(\l{})e_{k}^{22}\cr}}\right)\quad .\cr}}

\equation{eig}{\eqalign{
&\tau^{(1)} (\mu ) |\1l{1},\ldots , \1l{n_1}|G\rangle =
\nu^{(1)} (\mu ) |\1l{1},\ldots , \1l{n_1}|G\rangle = \cr
&\left(\prod_{i=1}^{n_1} {1\over \a{\mu}{\1l{i}}}\ \prod_{j=1}^n
\a{\mu}{\l{j}} \nu^{(2)}(\mu )
+\prod_{i=1}^{n_1} {1\over \a{\1l{i}}{\mu}}\right) |\1l{1},\ldots ,
\1l{n_1}|G\rangle \quad .}}

\equation{pbc1}{\prod_{j=1}^n{1\over a(\1l{k}-\l{j})}\
\prod_{\scr l=1\atop\scr l\neq k}^{n_1}
{\a{\1l{k}}{\1l{l}}\over\a{\1l{l}}{\1l{k}}}\ G^{b_{n_1}\ldots b_1} =
\tau^{(2)}(\1l{k})^{b_1\ldots b_{n_1}}_{a_1\ldots a_{n_1}}\
G^{a_{n_1}\ldots a_1}\ ,\ k=1,\ldots ,n_1\ .}

\equation{intnest3}{r^{(1)}(\l{}-\mu)\left(T_{n_1}^{(2)}(\l{})\otimes
T_{n_1}^{(2)}(\mu)\right)=\left(T_{n_1}^{(2)}(\mu)\otimes
T_{n_1}^{(2)}(\l{})\right) r^{(1)}(\l{}-\mu)\quad .}

\equation{intnest4}{\eqalign{D^{(2)}(\mu ) C^{(2)}(\la ) &= {1\over
\a{\mu}{\la} } C^{(2)}(\la )D^{(2)}(\mu ) + {\b{\la}{\mu}\over
\a{\la}{\mu}} C^{(2)}(\mu )D^{(2)}(\la )\cr
A^{(2)}(\mu ) C^{(2)}(\la ) &= {1\over
\a{\la}{\mu} } C^{(2)}(\la )A^{(2)}(\mu ) + {\b{\mu}{\la}\over
\a{\mu}{\la}} C^{(2)}(\mu )A^{(2)}(\la )\cr
C^{(2)}(\la)C^{(2)}(\mu) &= C^{(2)}(\mu)C^{(2)}(\la)\quad .\cr}}

\equation{vacnest2}{{\vac ^{(2)}_k = \pmatrix{0\cr 1\cr}\qquad ,\qquad
\vac ^{(2)} = \otimes_{k=1}^{n_1} \vac ^{(2)}_k}\quad .}

\equation{tvac2}{\eqalign{A^{(2)}(\mu)\vac ^{(2)} &= \prod_{j=1}^{n_1}
\a{\mu}{\l{j}}\vacnn\cr
D^{(2)}(\mu)\vac ^{(2)} &= \prod_{j=1}^{n_1}
\left(\a{\mu}{\l{j}}-\b{\mu}{\l{j}}\right)\vacnn = \prod_{j=1}^{n_1}
{\a{\mu}{\l{j}}\over \a{\l{j}}{\mu}}\vacnn\quad .\cr}}

\equation{state3}{|\2l{1},\ldots , \2l{n_2}\rangle =
C^{(2)}(\2l{1})\ C^{(2)}(\2l{2})\ldots C^{(2)}(\2l{n_2})
\ \vacnn\qquad .}

\equation{eig2}{\eqalign{
&\tau^{(2)} (\mu ) |\2l{1},\ldots , \2l{n_2}\rangle =
\nu^{(2)} (\mu ) |\2l{1},\ldots , \2l{n_2}\rangle = \cr
&-\left(\prod_{i=1}^{n_2} {1\over \a{\mu}{\2l{i}}}\prod_{j=1}^{n_1}
{\a{\mu}{\1l{j}}\over \a{\1l{j}}{\mu}}
+\prod_{i=1}^{n_2} {1\over \a{\2l{i}}{\mu}}
\prod_{j=1}^{n_1} \a{\mu}{\1l{j}}\right) |\2l{1},\ldots ,
\2l{n_2}\rangle \ ,}}

\equation{pbc2}{\prod_{i=1}^{n_1} \a{\1l{i}}{\2l{p}} = \prod_{\scr
j=1\atop\scr j\neq p}^{n_2}
{\a{\2l{j}}{\2l{p}}\over\a{\2l{p}}{\2l{j}}}\quad  ,\qquad p=1,\ldots
,n_2\ .}

\equation{nuFFBB}{\eqalign{\nu (\mu ) &=
(a(\mu ))^L\left(\prod_{j=1}^{N_e+N_l} {1\over
\a{\mu}{\l{j}}}\right) \nu ^{(1)}(\mu )\ + \left(
\prod_{j=1}^{N_e+N_l} {1\over \a{\l{j}}{\mu}}\right)\cr
\nu ^{(1)}(\mu ) &=
\prod_{i=1}^{N_e} {1\over \a{\mu}{\1l{i}}}
\left(\prod_{j=1}^{N_e+N_l} \a{\mu}{\l{j}}\right)
\nu^{(2)}(\mu) +\prod_{i=1}^{N_e} {1\over \a{\1l{i}}{\mu}}\cr
\nu^{(2)}(\mu) &= -\left(
\prod_{l=1}^{N_\down}{1\over a(\2l{l}-\mu)}
\prod_{j=1}^{N_e}\a{\mu}{\1l{j}}
+\prod_{l=1}^{N_\down}{1\over a(\mu -\2l{l})} \prod_{j=1}^{N_e}
{\a{\mu}{\1l{j}}\over a(\1l{j} - \mu)}\right)\quad .\cr}}

\equation{tildeFFBB}{\tl{k} = \l{k}+{i\over 2}\ ,\qquad
\tonel{j} = \1l{j}+i\ ,\qquad \ttwol{m} = \2l{m}+{i\over 2}\quad ,}

\equation{BAEFFBB}{\eqalign{
\left({\tl{k}-{i\over 2}\over\tl{k}+{i\over 2}}\right)^L &=
\prod_{j=1}^{N_e}\tb{\tl{k} -\tonel{j}}
\prod_{\scr l=1\atop\scr l\neq k}^{N_e+N_l}\tc{\tl{l}-\tl{k}}\ ,\quad
k=1,\ldots ,{N_e+N_l}\cr
\prod_{k=1}^{N_e+N_l}\tb{\tl{k} -\tonel{j}} &=
\prod_{m=1}^{N_\down}\tb{\ttwol{m} -\tonel{j}}
\quad ,\qquad j=1,\ldots ,N_e\cr
\prod_{\scr l=1\atop \scr l\neq m}^{N_\down}\tc{\ttwol{l} -\ttwol{m}}
&=\prod_{j=1}^{N_e}\ta{\ttwol{m} -\tonel{j}} \quad ,\qquad m=1,\ldots
,N_\down\ .\cr}}

\equation{Effbb}{\eqalign{E^0(\tl{1},\ldots ,\tl{n}) &= -i\ {\partial
\log(\tau (\mu))\over \partial\mu}\bigg |_{\mu=0}\cr
& = \sum_{j=1}^{N_e+N_l} {1\over{\tilde \l{j}}^2 +{1\over 4}} -L = -2
\sum_{j=1}^{N_e+N_l} cos(k_j) +2(N_e+N_l)-L\quad ,\cr}}

\equation{pFFBB}{p(\tl{1},\ldots ,\tl{n})=-i\sum_{j=1}^{N_e+N_l}
\log\left(\tb{\tl{j}}\right)=\sum_{j=1}^{N_e+N_l}k_j\quad .}


\equation{vacuum2}{{\vac _n = \pmatrix{0\cr 0\cr 0\cr 1\cr}\qquad
,\qquad \vac = \otimes_{n=1}^L \vac _n}\quad .}

\equation{Lop2}{L_n(\l{}) =
\left({\matrix{
a(\l{})+b(\l{})e_n^{11}&b(\l{})e_n^{21}&b(\l{})e_n^{31}&b(\la)e_n^{41}\cr
b(\l{})e_n^{12}&a(\l{})+b(\l{})e_n^{22}&b(\l{})e_n^{32}&b(\la)e_n^{42}\cr
b(\l{})e_n^{13}&b(\l{})e_n^{23}&a(\l{})-b(\la)e_n^{33}&-b(\la)e_n^{43}\cr
b(\l{})e_n^{14}&b(\l{})e_n^{24}&-b(\la)e_n^{34}&a(\la)-b(\la)e_n^{44}\cr
}}\right)\quad .}

\equation{tau22}{\tau (\mu) = str(T_L(\mu )) = A_{11}(\mu )+A_{22}(\mu
)- A_{33}(\mu)-D(\mu )\qquad .}

\equation{Tvac2}{T_L(\l{})\vac = \left(\matrix{(a(\l{}))^L&0&0&0\cr
0&(a(\l{}))^L&0&0\cr 0&0&(a(\l{}))^L&0\cr
C_1(\l{})&C_2(\l{})&C_3(\la)&\left({a(\la)\over
a(-\la)}\right)\cr}\right)\vac\qquad .}

\equation{state12}{|\l{1},\ldots , \l{n}|F\rangle = C_{a_1}(\l{1})\
C_{a_2}(\l{2})\ldots C_{a_n}(\l{n})\ \vac\  F^{a_n\ldots
a_1}\qquad ,}

\equation{AC2}{\eqalign{
A_{ab}(\mu)\ C_c(\l{}) &= (-1)^{\ge_a\ge_p+\ge_a+\ge_b}\
r_{BBF}(\mu-\la)^{dc}_{pb}\ C_p(\l{})\ A_{ad}(\mu)\cr
&\qquad +(-1)^{(\ge_a+1)(\ge_b+1)}\ {\b{\mu}{\la}\over\a{\mu}{\la}}
C_{b}(\mu) A_{ac}(\la )\cr
D(\mu)\ C_c(\l{}) &=
{1\over \a{\mu}{\la}}\ C_c(\l{})\ D(\mu)\ +\ {\b{\la}{\mu}\over
\a{\la}{\mu}} C_{c}(\mu ) D(\la )\cr
C_{a_1}(\l{1})\ C_{a_2}(\l{2}) &=
r_{FFB}(\l{1}-\l{2})^{b_1a_2}_{b_2a_1}\ C_{b_2}(\l{2})\
C_{b_1}(\l{1})\quad ,\cr}}

\equation{r2}{\eqalign{\srbbf{\mu}{a}{b}{c}{d}
&= b({\mu})I^{ab}_{cd} + a({\mu}){\pbbf}^{ab}_{cd}\quad ,\cr
\srffb{\mu}{a}{b}{c}{d}
&= b({\mu})I^{ab}_{cd} + a({\mu}){\pffb}^{ab}_{cd}\quad .\cr}}

\equation{Dstate2}{\eqalign{
D(\mu )|\l{1},\ldots , \l{n}|F\rangle &= \prod_{j=1}^n
{1\over \a{\mu}{\l{j}}} \left({a(\mu )\over a(-\mu )}\right)^L
|\l{1},\ldots , \l{n}|F\rangle\cr
&+ \sum_{k=1}^n\left({\tilde\Lambda}_k\right)^{b_1\ldots
b_n}_{a_1\ldots a_n} C_{b_k}(\mu ) \prod_{\scr j=1\atop \scr
j\neq k}^n C_{b_j}(\l{j})\vac F^{a_n\ldots a_1}\quad ,\cr}}

\equation{Astate2}{\eqalign{
&(A_{11}(\mu )+A_{22}(\mu )-A_{33}(\mu ))|\l{1},\ldots ,
\l{n}|F\rangle =\cr
&\qquad =\ (a(\mu ))^L\prod_{j=1}^n {1\over \a{\mu}{\l{j}}}
\prod_{l=1}^n C_{b_l}(\l{l})\vac\ \tau^{(1)}(\mu )^{b_1\ldots
b_n}_{a_1\ldots a_n}\ F^{a_n\ldots a_1} \cr
&\quad +\sum_{k=1}^n \left(\Lambda_k\right)^{b_1\ldots b_n}_{a_1\ldots a_n}
C_{b_k}(\mu )
\prod_{\scr j=1\atop \scr j\neq k}^n
C_{b_j}(\l{j})\vac F^{a_n\ldots a_1}\quad ,\cr}}

\equation{tausuth}{\eqalign{\tau^{(1)}(\mu )^{b_1\ldots
b_n}_{a_1\ldots a_n} &= (-1)^{\ge_c}\
\Le1{n}{\mu}^{cc_{n-1}}_{b_na_n}
\Le1{n-1}{\mu}^{c_{n-1}c_{n-2}}_{b_{n-1}a_{n-1}}
\ldots\Le1{1}{\mu}^{c_1c}_{b_1a_1}\times\cr
&\qquad\times (-1)^{\ge_c\sum_{i=1}^{n-1}(\ge_{b_i}+1) +
\sum_{i=1}^{n-1}\ge_{c_i}(\ge_{b_i}+1)}\quad .\cr}}

\equation{newgrtp}{\left(F\osl G\right)^{ab}_{cd} = F_{ab}G_{cd}\
(-1)^{(\ge_c +1)(\ge_a+\ge_b)} \quad .}

\equation{tau12}{\eqalign{\tau^{(1)}(\mu )^{b_1\ldots
b_n}_{a_1\ldots a_n} &= str( T_n^{(1)}(\mu ))=
str(\Le1{n}{\mu}\osl\Le1{n-1}{\mu}\osl\ldots
\osl\Le1{1}{\mu})\ .\cr}}

\equation{cancelbff}{\left((\Lambda_k)^{b_1\ldots b_n}_{a_1\ldots a_n} -
(\tilde\Lambda_k)^{b_1\ldots b_n}_{a_1\ldots a_n}\right)
F^{a_n\ldots a_1} = 0\quad }

\equation{pbcbff1}{F^{a_n\ldots a_1} = \left(a(-\l{k})\right)^L
\left(\tau^{(1)}(\l{k})F\right)^{a_n\ldots a_1}\ \ ,\ \ k=1,\ldots,
n\quad .}

\equation{lnest21}{L^{(1)}_k(\l{}) =
\left({\matrix{
a(\l{})+b(\l{})e_k^{11}&b(\l{})e_k^{21}&b(\l{})e_k^{31}\cr
b(\l{})e_k^{12}&a(\l{})+b(\l{})e_k^{22}&b(\l{})e_k^{32}\cr
b(\l{})e_k^{13}&b(\l{})e_k^{23}&a(\l{})-b(\la )e_k^{33}\cr}}
\right)\quad .}

\equation{otherr}{{{\hat r}(\mu )}^{ab}_{cd} = b(\mu ) \gd_{ab}\ \gd_{cd} +
a(\mu ) \gd_{ad}\ \gd_{bc}\ (-1)^{\ge_a +\ge_c + \ge_a\ge_c}\quad .}

\equation{vacnest22}{{\vac ^{(1)}_k = \pmatrix{0\cr 0\cr 1\cr}\qquad
,\qquad \vac ^{(1)} = \osl_{k=1}^n \vac ^{(1)}_k}\quad ,}

\equation{intnestbff1}{{\hat r}(\l{}-\mu)\
T_n^{(1)}(\l{})\osl T_n^{(1)}(\mu)=T_n^{(1)}(\mu)\osl
T_n^{(1)}(\l{})\ {\hat r}(\l{}-\mu)\quad .}

\equation{pbcbff2}{G^{a_{n_1}\ldots a_1} = \prod_{j=1}^n a(\l{j}-\1l{k})
\left(\tau^{(2)}(\1l{k})G\right)^{a_{n_1}\ldots a_1}\ \ ,\ \ k=1,\ldots,
{n_1}\quad ,}

\equation{tausuth2}{\eqalign{\tau^{(2)}(\mu )^{b_1\ldots
b_{n_1}}_{a_1\ldots a_{n_1}} &=
\Len{{n_1}}{\mu}^{cc_{{n_1}-1}}_{b_{n_1}a_{n_1}}
\Len{{n_1}-1}{\mu}^{c_{{n_1}-1}c_{{n_1}-2}}_{b_{{n_1}-1}a_{{n_1}-1}}
\ldots\Len{1}{\mu}^{c_1c}_{b_1a_1}\quad .\cr}}

\equation{nuBBFF}{\eqalign{\nu (\mu ) &=
(a(\mu ))^L\left(\prod_{j=1}^{N_b+N_\down} {1\over
\a{\mu}{\l{j}}}\right) \left(\nu ^{(1)}(\mu )\ - \left({1\over
a(-\mu)}\right)^L \right)\cr
\nu ^{(1)}(\mu ) &=
\prod_{i=1}^{N_b} {1\over \a{\mu}{\1l{i}}}
\left(\prod_{j=1}^{N_b+N_\down} \a{\mu}{\l{j}}\right)
\left(\nu^{(2)}(\mu) -\prod_{l=1}^{N_b+N_\down} {1\over
\a{\l{l}}{\mu}}\right) \cr
\nu ^{(2)}(\mu ) &=
\prod_{k=1}^{N_l}{1\over a(\mu - \2l{k})}
\prod_{j=1}^{N_b}\a{\mu}{\1l{j}}+
\prod_{j=1}^{N_l}{1\over \a{\2l{j}}{\mu}}\quad .\cr}}

\equation{tildeBBFF}{\tl{k} = \l{k}-{i\over 2}\ ,\qquad
\tonel{j} = \1l{j}-i\ ,\qquad \ttwol{m} = \2l{m}-{i\over 2}\quad ,}

\equation{BAEBBFF}{\eqalign{
\left({\tl{k}-{i\over 2}\over\tl{k}+{i\over 2}}\right)^L &=
\prod_{j=1}^{N_b}\tb{\tl{k} -\tonel{j}}
\prod_{\scr l=1\atop\scr l\neq k}^{N_b+N_\down}\tc{\tl{l}-\tl{k}}\ ,\
k=1,\ldots ,{N_b+N_\down}\cr
\prod_{k=1}^{N_\down+N_b}\tb{\tl{k} -\tonel{j}} &=
\prod_{m=1}^{N_l}\tb{\ttwol{m} -\tonel{j}}
\quad ,\quad\qquad j=1,\ldots ,N_b\cr
\prod_{\scr l=1\atop \scr l\neq m}^{N_l}\tc{\ttwol{l} -\ttwol{m}} &=
\prod_{j=1}^{N_b}\ta{\ttwol{m} -\tonel{j}} \quad ,\quad\qquad m=1,\ldots
,N_l\ .\cr}}

\equation{epBBFF}{\eqalign{E^0(\tl{1},\ldots ,\tl{n})
& = L-\sum_{j=1}^{N_b+N_\da} {1\over{\tilde \l{j}}^2 +{1\over 4}}  = -2
\sum_{j=1}^{N_b+N_\da} cos(k_j) +L-2(N_b+N_\da)\quad ,\cr
p(\tl{1},\ldots ,\tl{n}) &=-i\sum_{j=1}^{N_b+N_\da}
\log(\ta{\tl{j}})=\sum_{j=1}^{N_b+N_\da}k_j + (N_\up+1)\pi\quad ,\cr}}


\equation{nuFBBF}{\eqalign{\nu (\mu ) &=
(a(\mu ))^L\left(\prod_{j=1}^{N_\down+N_b} {1\over
\a{\mu}{\l{j}}}\right) \left(\nu ^{(1)}(\mu )\ - \left({1\over
a(-\mu)}\right)^L \right)\cr
\nu ^{(1)}(\mu ) &=
\prod_{j=1}^{N_\down+N_l} {1\over \a{\mu}{\1l{j}}}
\prod_{k=1}^{N_b+N_\down} \a{\mu}{\l{k}}\nu^{(2)}(\mu)
+\prod_{j=1}^{N_\down+N_l} {1\over
\a{\1l{j}}{\mu}}\cr
\nu ^{(2)}(\mu ) &=
\prod_{k=1}^{N_\down}{1\over a(\2l{k}-\mu)}
\left(1-\prod_{j=1}^{N_\down+N_l}\a{\mu}{\1l{j}}\right)
\quad .\cr}}

\equation{tildeFBBF}{\tl{k} = \l{k}-{i\over 2}\ ,\qquad
\ttwol{m} = \2l{m}+{i\over 2}\quad .}

\equation{BAEFBBF}{\eqalign{
\left({\tl{k}-{i\over 2}\over\tl{k}+{i\over 2}}\right)^L &=
\prod_{j=1}^{N_\down+N_l}\ta{\tl{k} -\1l{j}}\ ,\qquad k=1,\ldots
,{N_\down+N_b}\cr
\prod_{\scr l=1\atop \scr l\neq j}^{N_\down+N_l}\tc{\1l{j}
-\1l{l}} &=
\prod_{k=1}^{N_\down+N_b}\ta{\tl{k} -\1l{j}}\
\prod_{m=1}^{N_\down}\ta{\ttwol{m} -\1l{j}}\ ,\ j=1,\ldots
,N_\down+N_l\cr
1&=\prod_{j=1}^{N_\down+N_l}\ta{\ttwol{m} -\1l{j}} \ ,\qquad m=1,\ldots
,N_\down\ ,\cr}}


\equation{tildeBFFB}{\tl{k} = \l{k}+{i\over 2}\ ,\qquad
\ttwol{m} = \2l{m}-{i\over 2}\quad .}

\equation{BAEBFFB}{\eqalign{
\left({\tl{k}-{i\over 2}\over\tl{k}+{i\over 2}}\right)^L &=
\prod_{j=1}^{N_l+N_\down}\ta{\tl{k} -\1l{j}}\ ,\qquad k=1,\ldots
,{N_e+N_l}\cr
\prod_{\scr l=1\atop \scr l\neq j}^{N_l+N_\down}\tc{\1l{j}
-\1l{l}} &=
\prod_{k=1}^{N_e+N_l}\ta{\tl{k} -\1l{j}}\
\prod_{m=1}^{N_l}\ta{\ttwol{m} -\1l{j}}\ ,\quad j=1,\ldots
,N_l+N_\down\cr
1&=\prod_{m=1}^{N_l+N_\down}\ta{\ttwol{m} -\1l{j}} \quad ,\qquad
m=1,\ldots,N_l\ ,\cr}}

\equation{nuBFFB}{\eqalign{\nu (\mu ) &=
(a(\mu ))^L\left(\prod_{j=1}^{N_l+N_e} {1\over
\a{\mu}{\l{j}}}\right) \nu ^{(1)}(\mu )\ + \left(
\prod_{j=1}^{N_l+N_e} {1\over \a{\l{j}}{\mu}}\right)\cr
\nu ^{(1)}(\mu ) &=
\prod_{j=1}^{N_l+N_\down} {1\over \a{\mu}{\1l{j}}}
\left(\prod_{k=1}^{N_l+N_e} \a{\mu}{\l{k}}\right)
\left(\nu^{(2)}(\mu) -\prod_{l=1}^{N_l+N_e} {1\over
\a{\l{l}}{\mu}}\right) \cr
\nu ^{(2)}(\mu ) &=
\prod_{m=1}^{N_l}{1\over a(\mu -\2l{m})}
\prod_{j=1}^{N_l+N_\down} a(\mu -\1l{j})
\left(1-\prod_{p=1}^{N_l+N_\down}{1\over\a{\1l{p}}{\mu}}\right)
\quad .\cr}}


\equation{nuFBFB}{\eqalign{\nu (\mu ) &=
(a(\mu ))^L\left(\prod_{j=1}^{N_e+N_l} {1\over
\a{\mu}{\l{j}}}\right) \nu ^{(1)}(\mu )\ + \left(
\prod_{j=1}^{N_e+N_l} {1\over \a{\l{j}}{\mu}}\right)\cr
\nu ^{(1)}(\mu ) &=
\prod_{l=1}^{N_\down+N_l} {1\over \a{\mu}{\1l{l}}}
\left(\prod_{j=1}^{N_e+N_l} \a{\mu}{\l{j}}\right)
\left(\nu^{(2)}(\mu) -\prod_{k=1}^{N_e+N_l} {1\over
\a{\l{k}}{\mu}}\right) \cr
\nu ^{(2)}(\mu ) &=
\prod_{m=1}^{N_\down}{1\over a(\2l{m}-\mu)}
\left(1-\prod_{j=1}^{N_\down+N_l}\a{\mu}{\1l{j}}\right)
\quad .\cr}}

\equation{tildeFBFB}{\tl{k} = \l{k}+{i\over 2}\ ,\qquad
\ttwol{m} = \2l{m}+{i\over 2}\quad ,}

\equation{BAEFBFB}{\eqalign{
\left({\tl{k}-{i\over 2}\over\tl{k}+{i\over 2}}\right)^L &=
\prod_{j=1}^{N_\down+N_l}\ta{\tl{k} -\1l{j}}\quad ,\qquad k=1,\ldots
,{N_e+N_l}\cr
\prod_{k=1}^{N_e+N_l}\ta{\tl{k} -\1l{j}} &=
\prod_{m=1}^{N_\down}\ta{\ttwol{m} -\1l{j}}
\quad ,\quad\qquad j=1,\ldots ,N_\down+N_l\cr
1&=\prod_{j=1}^{N_\down+N_l}\ta{\ttwol{m} -\1l{j}} \quad ,\ \qquad
m=1,\ldots ,N_\down\ .\cr}}


{\equation{tildeBFBF}{\tl{k} = \l{k}-{i\over 2}\ ,\qquad
\ttwol{m} = \2l{m}-{i\over 2}\quad ,}

\equation{BAEBFBF}{\eqalign{
\left({\tl{k}-{i\over 2}\over\tl{k}+{i\over 2}}\right)^L &=
\prod_{j=1}^{N_\down+N_l}\ta{\tl{k} -\1l{j}}\quad ,\qquad k=1,\ldots
,{N_\down+N_b}\cr
\prod_{k=1}^{N_b+N_\da}\ta{\tl{k} -\1l{j}} &=
\prod_{m=1}^{N_l}\ta{\ttwol{m} -\1l{j}}
\quad ,\quad\qquad j=1,\ldots ,N_l+N_\down\cr
1&=\prod_{j=1}^{N_\down+N_l}\ta{\ttwol{m} -\1l{j}} \quad ,\ \qquad
m=1,\ldots ,N_l\ .\cr}}

\equation{nuBFBF}{\eqalign{\nu (\mu ) &=
(a(\mu ))^L\left(\prod_{j=1}^{N_b+N_\down} {1\over
\a{\mu}{\l{j}}}\right) \left(\nu ^{(1)}(\mu )\ - \left({1\over
a(-\mu)}\right)^L \right)\cr
\nu ^{(1)}(\mu ) &=
\prod_{i=1}^{N_l+N_\down} {1\over \a{\mu}{\1l{i}}}
\left(\prod_{j=1}^{N_b+N_\down} \a{\mu}{\l{j}}\right)
\nu^{(2)}(\mu) +\prod_{i=1}^{N_l+N_\down} {1\over \a{\1l{i}}{\mu}}\cr
\nu^{(2)}(\mu) &=
\prod_{k=1}^{N_l}{1\over a(\mu -\2l{k})}
\prod_{m=1}^{N_l+N_\down} a(\mu -\1l{m})
\left(1-\prod_{j=1}^{N_l+N_\down}{1\over\a{\1l{j}}{\mu}}\right)
\quad .\cr}}


\equation{boost}{B = \sum_n n H_{(2)}^{n,n+1}\qquad ,}

\equation{hcl2}{\eqalign{H_{(k+1)} &= i\lbrack B,H_{(k)}\rbrack \cr
& =i\lbrack {\tilde B},H_{(k)}\rbrack\qquad ,\cr}}

\equation{boost2}{{\tilde B} = -\sum_n n H^0_{n,n+1}\qquad .}

\equation{Rtilde}{{\tilde R}(\la ) = \Pg R(\la ) = b(\la )\Pg + a(\la
)I\quad ,}

\equation{rllkul}{{\tilde R}(\l{}-\mu)(L_n(\l{})\otimes
L_n(\mu)) = \left(I\otimes L_n(\mu)\right)\left(L_n(\l{})\otimes
I)\right){\tilde R}(\l{}-\mu)\quad .}

\equation{90rot}{\eqalign{{\tilde R}_{n,n+1}(\la -\mu )&\left(
L_{n,1}(\la )\otimes L_{n+1,1}(\mu )\right) =\cr
&\left(I_n\otimes L_{n+1,1}(\mu )\right)\left(L_{n,1}(\la )\otimes
I_{n+1}\right) {\tilde R}_{n,n+1}(\la -\mu )\quad ,\cr}}

\equation{90rot1}{\eqalign{&{\tilde R}_{n,n+1}(\la
-\mu)^{\ga_1\gc_1}_{\ga_2\gc_2}\ L_{n}(\la )^{a_1b_1}_{\gc_1\gb_1}\
L_{n+1}(\mu )^{b_1c_1}_{\gc_2\gb_2}\
(-1)^{\ge_{\gc_2}(\ge_{\gc_1}+\ge_{\gb_1})} = \cr
&L_{n+1}(\mu )^{a_1b_1}_{\ga_2\gc_2}\
L_{n}(\la )^{b_1c_1}_{\ga_1\gc_1}\
(-1)^{\ge_{\gc_2}(\ge_{\ga_1}+\ge_{\gc_1})}
{\tilde R}_{n,n+1}(\la -\mu )^{\gc_1\gb_1}_{\gc_2\gb_2}
\quad .\cr}}

\equation{90rot2}{{\tilde R}_{n,n+1}(\la -\mu )\ L_{n}(\la )\ L_{n+1}(\mu )
= L_{n+1}(\mu )\ L_{n}(\la )\ {\tilde R}_{n,n+1}(\la -\mu )\quad .}

\equation{diff}{\eqalign{
{\partial\over\partial\la}\left((\la +i) {\tilde R}_{n,n+1}(\la
)\right) &= I^{n,n+1}\cr
{\tilde R}_{n,n+1}(0) &= \Pg^{n,n+1}\quad .\cr}}

\equation{suth}{\left\lbrack \Pg^{n,n+1}, L_{n+1}(\mu )\otimes
L_{n}(\mu ) \right\rbrack = -i{\dot L}_{n+1}(\mu )\otimes L_{n}(\mu )
+i L_{n+1}(\mu )\otimes {\dot L}_{n}(\mu )\quad .}

\equation{suth2}{\left\lbrack H_{(2)}^{n,n+1}, L_{n+1}(\mu )\otimes
L_{n}(\mu ) \right\rbrack = i\left({\dot L}_{n+1}(\mu )\otimes
L_{n}(\mu ) - L_{n+1}(\mu )\otimes {\dot L}_{n}(\mu )\right)\quad .}

\equation{hcl-1}{\left\lbrack B,\tau (\mu )\right\rbrack = -i\ {\dot
\tau}(\mu )}

\equation{hcl0}{\left\lbrack B,\log(\tau \left(\mu )(\tau
(0))^{-1}\right)\right\rbrack =
-i{\partial\over\partial\mu} \log\left(\tau (\mu )\ (\tau
(0))^{-1}\right) - H_{(2)}\quad .}

\equation{Hagain}{H^0= -\sum_{j=1}^L H^0_{j,j+1}
= -\sum_{j=1}^L \Pg^{j,j+1}
= -\sum_{j=1}^L K^{\ga\gb}\ J_{j,\ga} \ J_{j+1,\gb}\qquad .}

\equation{H3}{\eqalign{H_{(3)} &= i\ \lbrack \tilde B, H_{(2)}\rbrack =
i\ \lbrack \tilde B, H^0\rbrack \cr
&=i\ \sum_{k=1}^L  \lbrack H^0_{k+1,k+2},
H^0_{k,k+1}\rbrack\cr
&=-i\ \sum_{k=1}^L K^{\ga\gb}\ K^{\gamma\gd}\ {f_{\gb\gamma}}^\ge
J_{k-1,\ga}\ J_{k,\ge}\ J_{k+1,\gd}\qquad .\cr}}

\equation{f}{ \lbrack J_{k,\ga}, J_{k,\gb}\rbrace := J_{k,\ga} J_{k,\gb} -
(-1)^{\epsilon_\ga\epsilon_\gb} J_{k,\gb} J_{k,\ga} =
{f_{\ga\gb}}^\gamma J_{k,\gamma}\qquad ,}

\equation{H4}{\eqalign{H_{(4)} &= i\lbrack \tilde B, H_{(3)}\rbrack
\cr
&=-2\sum_{k=1}^L K^{\mu\nu}\ K^{\ga\gb}\ K^{\gamma\gd}\
{f_{\gb\gamma}}^\ge\ {f_{\gd\mu}}^\omega J_{k-1,\ga}\ J_{k,\ge}\
J_{k+1,\omega}\ J_{k+2,\nu}\cr
&\qquad +\sum_{k=1}^L { P}^{k-1,k+1}\ -2\sum_{k=1}^L \Pg^{k,k+1}
\qquad ,\cr}}

\equation{P}{{P}^{k-1,k+1} =
\Pg^{k-1,k}\Pg^{k,k+1}\Pg^{k-1,k}\quad .}


\equation{baefbbf3}{1=\prod_{j=1}^{N_\down+N_l}\ta{\ttwol{m} -\1l{j}}
\quad ,\ \qquad m=1,\ldots ,N_\da\ .}

\equation{pol}{p(w)=\prod_{j=1}^{N_l+N_{\down}} (w-\1l{j}-{i\over
2}) - \prod_{j=1}^{N_l+N_{\down}} (w-\1l{j}+{i\over 2}) = 0\quad
.}

\equation{res}{\sum_{j=1}^{N_\da} -i\log\left({\1l{l}-w_j+{i\over
2}\over\1l{l}-w_j-{i\over 2}}\right) =
\sum_{j=1}^{N_\da}{1\over 2\pi i}\oint_{C_j} dz
(-i)\log\left({\1l{l}-z+{i\over 2}\over\1l{l}-z-{i\over
2}}\right)\ {d\over dz}\log(p(z))\ ,}

\equation{res2}{\sum_{j=1}^{N_\da}
-i\log\left({\1l{l}-w_j+{i\over
2}\over\1l{l}-w_j-{i\over 2}}\right) =  -\sum_{j=1}^{N_l}
-i\log\left({\1l{l}-{w^\prime}_j+{i\over
2}\over\1l{l}-{w^\prime}_j-{i\over
2}}\right)-i\log\left({p(z_n)\over p(z_p)}\right)\quad ,}

\equation{cut}{\eqalign{p(z_n) &= -
\prod_{m=1}^{N_l+N_\downarrow}\left(\1l{l}-\1l{m}+i \right)\cr
p(z_p) &= \prod_{m=1}^{N_l+N_\downarrow}\left(\1l{l}-\1l{m}-i
\right)\quad .\cr}}

\equation{id}{\prod_{j=1}^{N_\da}{\1l{l}-w_j+{i\over 2}\over
\1l{l}-w_j-{i\over 2}}
=\prod_{k=1}^{N_l}{\1l{l}-{w^\prime}_k-{i\over 2}\over
\1l{l}-{w^\prime}_k+{i\over 2}}\  \prod_{\scr m=1\atop \scr m\neq
l}^{N_l+N_\downarrow} {\tc{\1l{l} - \1l{m}}}\quad .}

\equation{baebfbf2}{\prod_{k=1}^{N_b+N_\da}\ta{\tl{k} -\1l{l}} =
\prod_{m=1}^{N_l}\ta{{w^\prime}_{m}-\1l{l}}
\quad ,\quad\qquad j=1,\ldots ,N_l+N_\down \ .}

\equation{baebfbf1}{\left({\tl{k}-{i\over 2}\over\tl{k}+{i\over
2}}\right)^L = \prod_{j=1}^{N_\down+N_l}\ta{\tl{k} -\1l{j}}\quad
,\qquad k=1,\ldots ,{N_\down+N_b},}

\equation{pol2}{p(w)=\left(w-{i\over 2}\right)^L \prod_{j=1}^{N_l+N_{\down}}
(w-\1l{j}+{i\over 2}) - \left(w+{i\over 2}\right)^L
\prod_{j=1}^{N_l+N_{\down}} (w-\1l{j}-{i\over 2}) = 0\quad .}

\equation{id2}{
\prod_{k=1}^{N_b+N_\da}\tb{\1l{l}-w_k} =
\prod_{j=1}^{L-N_h}\ta{\1l{l}-w^\prime_j}\
\prod_{\scr m=1\atop \scr m\neq
l}^{N_l+N_\da} {\tc{\1l{m} - \1l{l}}}\quad .}

\equation{baebffb}{\prod_{m=1}^{N_l}\ta{\ttwol{m}-\1l{j}}=
\prod_{l=1}^{N_e+N_l}\ta{\tonel{j} -w^\prime_j}\
\prod_{\scr k=1\atop \scr k\neq
j}^{N_l+N_\da} {\tc{\1l{k} - \1l{j}}}
\ ,\ j=1,\ldots ,N_l+N_\da .}

\equation{pol3}{\eqalign{p(w)&=
\prod_{k=1}^{N_b+N_\da}(\tl{k}-w-{i\over 2})\
\prod_{j=1}^{N_l}({w-\ttwol{j}}-{i\over 2})
-\ \prod_{k=1}^{N_b+N_\da}({\tl{k}-w}+{i\over 2})\ \prod_{j=1}^{N_l}
({w-\ttwol{j}}+{i\over 2})\cr
& =0\quad .\cr}}

\equation{id3}{\eqalign{
\prod_{j=1}^{N_l+N_\da}\ta{\tl{k}-w_j} &= \prod_{\scr
l=1\atop\scr l\neq k}^{N_b+N_\da}\tc{\tl{l}-\tl{k}}\
\prod_{j=1}^{N_b}\tb{\tl{k}-w^\prime_j}\ ,\quad k=1\ldots N_b+N_\da\cr
\prod_{j=1}^{N_l+N_\da}\ta{\ttwol{m}-w_j} &= \prod_{\scr
l=1\atop\scr l\neq m}^{N_l}\tc{\ttwol{l}-\ttwol{m}}\
\prod_{j=1}^{N_b}\tb{\ttwol{m}-w^\prime_j}\ ,\quad m=1\ldots N_l\
.\cr}}


\equation{front}{\eqalign{\prod_{i=1}^n C_{a_i} (\l{i}) &= C_{b_k}
(\l{k})\  \prod_{i=1}^{k-1} C_{b_i}(\l{i})\ \prod_{j=k+1}^n C_{a_j}
(\l{j})\ S(\l{k})^{b_1\ldots b_k}_{a_1\ldots a_k}\quad
,\cr\noalign{\vskip 4pt}
S(\l{k})^{b_1\ldots b_k}_{a_1\ldots
a_k}&=\r{\l{k-1}}{\l{k}}{b_{k-1}}{a_k}{c_{k-1}}{a_{k-1}}\
\r{\l{k-2}}{\l{k}}{b_{k-2}}{c_{k-1}}{c_{k-2}}{a_{k-2}}\ldots
\r{\l{1}}{\l{k}}{b_{1}}{c_2}{b_{k}}{a_{1}}\quad .\cr}}

\equation{laktilde}{\left({\tilde\Lambda}_k F\right)^{b_1\ldots b_n}
= S(\l{k})^{b_1\ldots b_k}_{a_1\ldots a_k}\ F^{b_n\ldots
b_{k+1}a_k\ldots a_1}
\left(-{\b{\l{k}}{\mu}\over\a{\l{k}}{\mu}}\right)\prod_{\scr
i=1\atop\scr i\neq k}^n {1\over \a{\l{i}}{\l{k}}}
\quad .}

\equation{lak1}{\eqalign{
\left({\Lambda}_{k,1} F\right)^{b_1\ldots b_n}
&=- S(\l{k})^{c_1\ldots c_k}_{a_1\ldots a_k}\ F^{a_n\ldots a_1}
\left({\b{\mu}{\l{k}}\over\a{\mu}{\l{k}}}\right) \gd_{b_k,1}\
\prod_{\scr
i=1\atop\scr i\neq k}^n {1\over \a{\l{k}}{\l{i}}}
\cr
&\times \
\r{\l{k}}{\l{1}}{d_1}{c_1}{b_1}{c_k}
\r{\l{k}}{\l{2}}{d_2}{c_2}{b_2}{d_1}\ldots
\r{\l{k}}{\l{k-1}}{d_{k-1}}{c_{k-1}}{b_{k-1}}{d_{k-2}}\cr
&\times \
\r{\l{k}}{\l{k+1}}{d_k}{a_{k+1}}{b_{k+1}}{d_{k-1}}
\r{\l{k}}{\l{k+2}}{d_{k+1}}{a_{k+2}}{b_{k+2}}{d_{k}}\ldots
\r{\l{k}}{\l{n}}{d_{n-1}}{a_{n}}{b_{n}}{d_{n-2}}\cr
&\times (a(\l{k}))^L \ \gd_{d_{n-1},1}\ (-1)^{\sum_{
i=1\atop i\neq k}^n \ge_{b_i}\ge_1}\quad .\cr}}

\equation{invr}{\r{\l{1}}{\l{2}}{b_1}{a_1}{b_2}{a_2}\
\r{\l{2}}{\l{1}}{c_1}{b_1}{c_2}{b_2} = I^{a_1 c_1}_{a_2 c_2} = \gd
_{a_1 c_1}\ \gd _{a_2 c_2}\quad ,}

\equation{lak2}{\eqalign{
\left(\Lambda_k F\right)^{b_1\ldots b_n}
&= -S(\l{k})^{c_1\ldots c_k}_{a_1\ldots a_k}\ F^{a_n\ldots a_1}
\left({\b{\mu}{\l{k}}\over\a{\mu}{\l{k}}}\right)\prod_{\scr
i=1\atop\scr i\neq k}^n {1\over \a{\l{k}}{\l{i}}}
\ (a(\l{k}))^L \cr
&\times \
\r{\l{k}}{\l{1}}{d_1}{c_1}{b_1}{c_k}
\r{\l{k}}{\l{2}}{d_2}{c_2}{b_2}{d_1}\ldots
\r{\l{k}}{\l{k-1}}{d_{k-1}}{c_{k-1}}{b_{k-1}}{d_{k-2}}\
(-1)^{\sum_{ i=1\atop i\neq k}^n \ge_{b_i}\ge_{b_k}}\
\cr
&\times \
\r{\l{k}}{\l{k+1}}{d_k}{a_{k+1}}{b_{k+1}}{d_{k-1}}
\r{\l{k}}{\l{k+2}}{d_{k+1}}{a_{k+2}}{b_{k+2}}{d_{k}}\ldots
\r{\l{k}}{\l{n}}{b_{k}}{a_{n}}{b_{n}}{d_{n-2}}\quad .\cr}}

\equation{raise}{r(\la )^{ab}_{cd} = (r(\la )\pg)^{ad}_{cb}\
(-1)^{\ge_a\ge_c} = (-1)^{\ge_a\ge_c} L^{(1)}(\la )^{ad}_{cb}\quad .}

\equation{lak3}{\eqalign{
&\left({\Lambda}_k F\right)^{b_1\ldots b_n}
= {\b{\mu}{\l{k}}\over\a{\mu}{\l{k}}}\prod_{\scr
i=1\atop\scr i\neq k}^n {1\over \a{\l{k}}{\l{i}}}
\ (a(\l{k}))^L \ F^{a_n\ldots a_kb_{k-1}\ldots b_1}\cr
&\times\
\LE{{n}}{\l{k}}{b_{k}}{d_{n-2}}{b_{n}}{a_{n}}
\LE{{n-1}}{\l{k}}{d_{n-2}}{d_{n-3}}{b_{n-1}}{a_{n-1}}\ldots
\LE{{k+1}}{\l{k}}{d_k}{a_{k}}{b_{k+1}}{a_{k+1}}\ \cr
&\times\ (-1)^{\sum_{ i=1\atop i\neq k}^n \ge_{b_i}\ge_{b_k} +
\sum_{j=k}^{n-2}\ge_{d_j}\ge_{b_{j+1}}+\ge_{b_k}\ge_{b_n} }
\ \ .\cr}}

\equation{Sr}{S(\l{k})^{c_1\ldots c_k}_{a_1\ldots a_k}\
\r{\l{k}}{\l{1}}{d_1}{c_1}{b_1}{c_k}\ldots
\r{\l{k}}{\l{k-1}}{d_{k-1}}{c_{k-1}}{b_{k-1}}{d_{k-2}} =
\prod_{i=1}^{k-1}\gd_{a_i,b_i}\ \gd_{d_{k-1},a_k}\quad.}


\equation{unwn1a}{\eqalign{
&\left({\Lambda}^{(1)}_k G\right)^{g_1\ldots g_{n_1}}\hskip -1pt
= {\b{\mu}{\1l{k}}\over\a{\mu}{\1l{k}}}\prod_{\scr
i=1\atop\scr i\neq k}^{n_1} {1\over \a{\1l{k}}{\1l{i}}}
\prod_{j=1}^n a(\1l{k}-\l{j})
\ G^{f_{n_1}\ldots f_k g_{k-1}\ldots g_1} (-1)^{k+1}
\cr
&\times\
\LEn{{n}}{\1l{k}}{g_{k}}{d_{n-2}}{g_{n}}{f_{n}}
\LEn{{n-1}}{\1l{k}}{d_{n-2}}{d_{n-3}}{g_{n-1}}{f_{n-1}}\ldots
\LEn{{k+1}}{\1l{k}}{d_k}{f_{k}}{g_{k+1}}{f_{k+1}}
\ \ .\cr}}

\equation{unwn1b}{\left({\tilde\Lambda}^{(1)}_k G\right)^{g_1\ldots
g_{n_1}}
= S^{(1)}(\1l{k})^{g_1\ldots g_k}_{f_1\ldots f_k}\ F^{g_{n_1}\ldots
g_{k+1}f_k\ldots f_1}
\left(-{\b{\1l{k}}{\mu}\over\a{\1l{k}}{\mu}}\right)\prod_{\scr
i=1\atop\scr i\neq k}^{n_1} {1\over \a{\1l{i}}{\1l{k}}}
\ ,}

\equation{sone}{\eqalign{&S^{(1)}(\1l{k})^{g_1\ldots g_k}_{f_1\ldots
f_k}=\cr
&\qquad \rn{\1l{k-1}}{\1l{k}}{g_{k-1}}{f_k}{c_{k-1}}{f_{k-1}}\ \
\rn{\1l{k-2}}{\1l{k}}{g_{k-2}}{c_{k-1}}{c_{k-2}}{f_{k-2}}\ \ldots\
\rn{\1l{1}}{\1l{k}}{g_{1}}{c_2}{g_{k}}{f_{1}}.\cr}}


\equation{unwn2}{\eqalign{
\Lambda^{(2)}_k &= {\b{\mu}{\2l{k}}\over \a{\mu}{\2l{k}}} \prod_{\scr
j=1\atop\scr j\neq k}^{n_2} {1\over
\a{\2l{j}}{\2l{k}}}\prod_{l=1}^{n_1}\a{\2l{k}}{\1l{l}}\cr
{\tilde\Lambda}^{(2)}_k &= {\b{\2l{k}}{\mu}\over
\a{\2l{k}}{\mu}} \prod_{\scr j=1\atop\scr j\neq
k}^{n_2} {1\over
\a{\2l{k}}{\2l{j}}}\prod_{l=1}^{n_1}{\a{\2l{k}}{\1l{l}}\over
\a{\1l{l}}{\2l{k}}}\quad .\cr}}


\equation{front2}{\eqalign{\prod_{i=1}^n C_{a_i} (\l{i}) &= C_{e_k} (\l{k})\
\prod_{i=1}^{k-1} C_{e_i}(\l{i})\ \prod_{j=k+1}^n C_{a_j}
(\l{j})\ S(\l{k})^{e_1\ldots e_k}_{a_1\ldots a_k}\quad
,\cr\noalign{\vskip 4pt}
S(\l{k})^{e_1\ldots e_k}_{a_1\ldots
a_k}&=\cr
&\hskip -30pt\rffb{\l{k}}{\l{k-1}}{e_{k-1}}{a_k}{c_{k-1}}{a_{k-1}}\
\rffb{\l{k}}{\l{k-2}}{e_{k-2}}{c_{k-1}}{c_{k-2}}{a_{k-2}}\ldots
\rffb{\l{k}}{\l{1}}{e_{1}}{c_2}{e_{k}}{a_{1}}\quad .\cr}}

\equation{laktilde2}{\left({\tilde\Lambda}_k F\right)^{b_1\ldots b_n}
= S(\l{k})^{b_1\ldots b_k}_{a_1\ldots a_k}\ F^{b_n\ldots
b_{k+1}a_k\ldots a_1}
\left({\b{\l{k}}{\mu}\over\a{\l{k}}{\mu}}\right)\prod_{\scr
i=1\atop\scr i\neq k}^n {1\over \a{\l{k}}{\l{i}}} \left({a(\l{k})\over
a(-\l{k})} \right)\quad .}

\equation{lak22}{\eqalign{
\left(\Lambda_k F\right)^{b_1\ldots b_n}
&= S(\l{k})^{e_1\ldots e_k}_{a_1\ldots a_k}\ F^{a_n\ldots a_1}
\left({\b{\mu}{\l{k}}\over\a{\mu}{\l{k}}}\right)
\prod_{\scr i=1\atop\scr i\neq k}^n {1\over \a{\l{k}}{\l{i}}}
\ (a(\l{k}))^L \ \cr
&\times \
\r{\l{k}}{\l{1}}{d_1}{e_1}{b_1}{e_k}
\r{\l{k}}{\l{2}}{d_2}{e_2}{b_2}{d_1}\ldots
\r{\l{k}}{\l{k-1}}{d_{k-1}}{e_{k-1}}{b_{k-1}}{d_{k-2}}\
\cr
&\times \
\r{\l{k}}{\l{k+1}}{d_k}{a_{k+1}}{b_{k+1}}{d_{k-1}}
\r{\l{k}}{\l{k+2}}{d_{k+1}}{a_{k+2}}{b_{k+2}}{d_{k}}\ldots
\r{\l{k}}{\l{n}}{b_{k}}{a_{n}}{b_{n}}{d_{n-2}}\ \cr
&\times \
(-1)^{\ge_{e_k}+\sum_{i=1}^{n-2} \ge_{d_i}+\ge_{b_k}\sum_{i=1\atop
i\neq k}^n(\ge_{b_i}+1)}(-1)^{(\ge_{b_k}+1)(\ge_{b_k}+1)}
\quad .\cr}}

\equation{invr2}{r_{FFB}(\la)^{e_1c_2}_{e_ka_1}\
r_{BBF}(\la)^{d_1e_1}_{b_1e_k}\ (-1)^{\ge_{e_k}} =
(-1)^{\ge_{b_1}} (b(\la)-a(\la)) I^{d_1c_2}_{b_1a_1}\quad .}


\equation{lws}{\eqalign{
0 & =\ \eta |\l{1},\ldots , \l{n}|F\rangle \cr
  & =\ Q_1  |\l{1},\ldots , \l{n}|F\rangle \cr
  & =\ Q_{-1}  |\l{1},\ldots , \l{n}|F\rangle \cr}
\eqalign{  & =\ S |\l{1},\ldots , \l{n}|F\rangle \cr
  & =\ {\tilde Q}^\dagger_1  |\l{1},\ldots , \l{n}|F\rangle \cr
  & =\ {\tilde Q}^\dagger_{-1}  |\l{1},\ldots , \l{n}|F\rangle
\quad .\cr}}

\equation{lowop}{\eqalign{
S &=\sum_{k=1}^L e_k^{21}\quad ,\quad
\eta =\sum_{k=1}^L e_k^{43}\quad ,\quad
Q_1 =\sum_{k=1}^L e_k^{42}\cr
Q_{-1} &=\sum_{k=1}^L e_k^{41}\quad ,\quad
{\tilde Q}^\dagger_1 =-\sum_{k=1}^L e_k^{31}\quad ,\quad
{\tilde Q}^\dagger_{-1} =\sum_{k=1}^L e_k^{32}
\quad ,\cr}}

\equation{esubk}{\left(e_k^{ij}\right)^{\ga\gb} = \gd_{i\ga}\
\gd_{j\gb}\quad .}

\equation{e}{\left(e^{ij}\right)^{ab} = \gd_{ia}\
\gd_{jb}\quad .}

\equation{lgen}{\eqalign{
\left\lbrack \left(L_k(\mu)\right)^{ab} ,
e_k^{ij}\right\rbrace_{quantum} &= (-1)^{(\ge_i+\ge_j)\ge_b}\ \left(
\left(e^{ij}\ L_k(\mu)\right)^{ab} - \left(L_k(\mu)\
e^{ij}\right)^{ab}\right)\cr
&=(-1)^{(\ge_i+\ge_j)\ge_b}\ \left(\left\lbrack e^{ij} , L_k(\mu)
\right\rbrack_{matrix}\right)^{ab} \quad.\cr}}

\equation{tgen}{\eqalign{\left\lbrack \left(T_L(\mu)\right)^{ab} ,
e_k^{ij}\right\rbrace_{quantum} &= (-1)^{(\ge_i+\ge_j)\ge_b}\
L_L(\mu)^{ac_n}\ldots L_{k+1}(\mu)^{c_{k+2}c_{k+1}}\ \times\cr
&\qquad\left(\left(e^{ij}\ L_k(\mu)\right)^{c_{k+1}c_k} - \left(L_k(\mu)\
e^{ij}\right)^{c_{k+1}c_k}\right)\ \times\cr
&\qquad L_{k-1}(\mu)^{c_{k}c_{k-1}}\ldots L_{1}(\mu)^{c_{2}b}\quad .}}

\equation{tgen2}{\left\lbrack \sum_{k=1}^Le_k^{ij} ,
\left(T_L(\mu)\right)^{ab} \right\rbrace_{quantum} =
(-1)^{(\ge_i+\ge_j)\ge_b}\ \left(\left\lbrack e^{ij} , T_L(\mu)
\right\rbrack_{matrix}\right)^{ab} .}

\equation{Cgen}{\eqalign{
\left\lbrack \eta , C_a(\la)\right\rbrack &=
\gd_{a3} D(\la)-A_{3a}(\la)\cr
\left\lbrack Q_1 , C_a(\la)\right\rbrace &=
\gd_{a2} D(\la)-A_{2a}(\la)\cr
\left\lbrack Q_{-1} , C_a(\la)\right\rbrace &=
\gd_{a1} D(\la)-A_{1a}(\la)\cr
\left\lbrack -\tqd_1 , C_a(\la)\right\rbrace &=
\gd_{a1} C_{3}(\la)\cr
\left\lbrack \tqd_{-1} , C_a(\la)\right\rbrace &=
\gd_{a2} C_{3}(\la)\cr
\left\lbrack S, C_a(\la)\right\rbrack &= \gd_{a1} C_2(\la)
\quad .\cr}}

\equation{annihilate}{0=\eta\vac = \tqd_1\vac = \tqd_{-1}\vac =
Q_1\vac = Q_{-1}\vac = S\vac  \quad .}

\equation{lw1}{\eqalign{
J_a\ &|\l{1},\ldots , \l{n}|F\rangle \ = \cr
&= \left\lbrack J_a , \prod_{i=1}^n C_{a_i}(\l{i})\right\rbrace\vac\
F^{a_n\ldots a_1}\cr
&= \sum_{k=1}^n \prod_{i=1}^{k-1} C_{a_i}(\l{i})
{\left\lbrack J_a , C_{a_k}(\l{k})\right\rbrace}
\prod_{j=k+1}^{n} C_{a_j}(\l{j})\vac\ F^{a_n\ldots a_1}
(-1)^{\ge_a\sum_{l=1}^{k-1}\ge_{a_l}}\cr
&= \sum_{k=1}^n \prod_{i=1}^{k-1} C_{a_i}(\l{i})
{\left(\gd_{aa_k} D(\l{k})-A_{aa_k}(\l{k})\right)}
\prod_{j=k+1}^{n} C_{a_j}(\l{j})\vac\ F^{a_n\ldots a_1}
(-1)^{\ge_a\sum_{l=1}^{k-1}\ge_{a_l}}\cr
&= \sum_{k=1}^n \prod_{i=1}^{k-1} C_{b_i}(\l{i})
\prod_{j=k+1}^{n} C_{b_{j}}(\l{j})\vac
\left({\Omega_k}\right)^{b_1\ldots b_{k-1}ab_{k+1}\ldots b_{n}}
(-1)^{\ge_a\sum_{l=1}^{k-1}\ge_{a_l}}
\quad .\cr}}

\equation{lw2}{
{\left(\gd_{aa_1} D(\l{1})-A_{aa_1}(\l{1})\right)}
\prod_{j=2}^{n} C_{a_j}(\l{j})\vac\ F^{a_n\ldots a_1}}

\equation{o1}{\eqalign{\left(\Omega_1\right)^{ab_2\ldots
b_{n}} =
&\gd_{aa_1}\left\lbrack\prod_{i=2}^n {1\over\a{\l{i}}{\l{1}}}
\prod_{j=2}^{n}\gd_{a_jb_{j}} \right\rbrack\ \ F^{a_n\ldots a_1}\cr
& - \prod_{i=2}^n {1\over \a{\l{1}}{\l{i}}}\
\r{\l{1}}{\l{2}}{d_1}{a_{2}}{b_{2}}{a_1}\
\r{\l{1}}{\l{3}}{d_{2}}{a_{3}}{b_{3}}{d_{1}}\ldots\
\r{\l{1}}{\l{n}}{a}{a_{n}}{b_{n}}{d_{n-2}}\cr
&\quad\times\ (-1)^{\sum_{i=2}^n \ge_{b_i}\ge_a}(a(\l{1}))^L\
F^{a_n\ldots a_1} \quad .\cr}}

\equation{kfront}{\eqalign{\prod_{i=1}^n C_{a_i} (\l{i})
\vac F^{a_n\ldots a_1} &=
C_{c_k}(\l{k})\  \prod_{i=1}^{k-1} C_{c_i}(\l{i})\ \prod_{j=k+1}^n
C_{a_j}(\l{j})\ S(\l{k})^{c_1\ldots c_k}_{a_1\ldots a_k}\vac
F^{a_n\ldots a_1} \quad ,\cr\noalign{\vskip 4pt}
S(\l{k})^{c_1\ldots c_k}_{a_1\ldots
a_k}&=\r{\l{k-1}}{\l{k}}{c_{k-1}}{a_k}{d_{k-1}}{a_{k-1}}\
\r{\l{k-2}}{\l{k}}{c_{k-2}}{d_{k-1}}{d_{k-2}}{a_{k-2}}\ldots
\r{\l{1}}{\l{k}}{c_{1}}{d_2}{c_{k}}{a_{1}}\ .\cr}}

\equation{lw3}{\eqalign{
J_a\ &|\l{1},\ldots , \l{n}|F\rangle \ = \cr
&= \left\lbrack J_a \ ,\
C_{b_k}(\l{k})\  \prod_{i=1}^{k-1} C_{b_i}(\l{i})\ \prod_{j=k+1}^n
C_{a_j}(\l{j})\ S(\l{k})^{b_1\ldots b_k}_{a_1\ldots a_k}
\right\rbrace\vac\ F^{a_n\ldots a_1}\quad .\cr}}

\equation{ok}{\eqalign{\left(\Omega_k\right)^{b_1\ldots
b_{k-1}ab_{k+1} \ldots b_{n}}& =
\left\lbrack\prod_{\scr i=1\atop\scr i\neq k}^n
{1\over\a{\l{i}}{\l{k}}}
\prod_{j=k+1}^{n}\gd_{a_jb_{j}} \
S(\l{k})^{b_1\ldots b_{k-1}a}_{a_1\ldots a_k}\right\rbrack\ \
F^{a_n\ldots a_1}\cr
&\hskip -18pt - \prod_{\scr i=1\atop\scr i\neq k}^n {1\over
\a{\l{k}}{\l{i}}}\ S(\l{k})^{c_1\ldots c_k}_{a_1\ldots a_k}\
(-1)^{\sum_{ i=1\atop i\neq k}^n \ge_{b_i}\ge_a}\ F^{a_n\ldots a_1}\cr
&\hskip -18pt\qquad\times\r{\l{k}}{\l{1}}{d_1}{c_1}{b_1}{c_k}
\r{\l{k}}{\l{2}}{d_2}{c_2}{b_2}{d_1}\ldots
\r{\l{k}}{\l{k-1}}{d_{k-1}}{c_{k-1}}{b_{k-1}}{d_{k-2}}\cr
&\hskip -18pt\qquad\times \
\r{\l{k}}{\l{k+1}}{d_k}{a_{k+1}}{b_{k+1}}{d_{k-1}}
\r{\l{k}}{\l{k+2}}{d_{k+1}}{a_{k+2}}{b_{k+2}}{d_{k}}\ldots
\r{\l{k}}{\l{n}}{a}{a_{n}}{b_{n}}{d_{n-2}}\cr
&\hskip -18pt\qquad\times (a(\l{k}))^L \quad .\cr}}

\equation{lim}{\eqalign{
\tqd_{-1} = \sum_{k=1}^L e_k^{32} = -i\ \lim_{\mu\rightarrow\infty}
\mu\ A_{23}(\mu)\cr
\tqd_{1} = -\sum_{k=1}^L e_k^{31} = i\ \lim_{\mu\rightarrow\infty}
\mu\ A_{13}(\mu)\quad .\cr}}

\equation{lim2}{-i\ \lim_{\mu\rightarrow\infty}\mu\ A_{ab}(\mu)
|\l{1}\ldots\l{n}|F\rangle = -i\prod_{l=1}^nC_{b_l}(\l{l})\vac\
\lim_{\mu\rightarrow\infty}\mu\left({T^{(1)}_n(\mu)}^{ab}\right)^{b_1\ldots
b_n}_{a_1\ldots a_n} \ F^{a_1\ldots a_n}\quad .}

\equation{lim3}{\eqalign{
\tqd_1\ |\l{1}\ldots\l{n}|F\rangle = \prod_{l=1}^n C_{b_l}(\l{l})\vac\
\lim_{\mu\rightarrow\infty}i\ \mu
\left(B_1^{(1)}(\mu)F\right)^{b_1\ldots b_n}\cr
\tqd_{-1}\ |\l{1}\ldots\l{n}|F\rangle = \prod_{l=1}^n C_{b_l}(\l{l})\vac\
\lim_{\mu\rightarrow\infty}-i\ \mu
\left(B_2^{(1)}(\mu)F\right)^{b_1\ldots b_n}\quad .}}

\equation{j1}{J_a^{(1)} = \lim_{\mu\rightarrow\infty}-i\ \mu\
B^{(1)}_a(\mu)\ ,\quad a=1,2\quad .}

\equation{j1f}{J^{(1)}_a\ F = 0\ ,\quad a=1,2\qquad .}

\equation{jc1}{\left\lbrack J^{(1)}_a, C^{(1)}_b(\1l{})\right\rbrack
= \gd_{ab} D^{(1)}(\1l{})-A^{(1)}_{ab}(\1l{})\quad .}

\equation{nlw1}{\eqalign{
J^{(1)}_a\ &|\1l{1},\ldots , \1l{n_1}|G\rangle \ = \cr
&= \left\lbrack J^{(1)}_a , \prod_{i=1}^{n_1}
C^{(1)}_{a_i}(\1l{i})\right\rbrace\vacn\ G^{a_{n_1}\ldots a_1}\cr
&= \sum_{k=1}^{n_1} \prod_{i=1}^{k-1} C^{(1)}_{a_i}(\1l{i})
{\left\lbrack J^{(1)}_a , C^{(1)}_{a_k}(\1l{k})\right\rbrace}
\prod_{j=k+1}^{n_1} C^{(1)}_{a_j}(\1l{j})\vacn\ G^{a_{n_1}\ldots a_1}
(-1)^{\ge_a\sum_{l=1}^{k-1}\ge_{a_l}}\cr
&= \sum_{k=1}^{n_1} \prod_{i=1}^{k-1} C^{(1)}_{a_i}(\1l{i})
{\left(\gd_{aa_k} D^{(1)}(\1l{k})-A^{(1)}_{aa_k}(\1l{k})\right)}
\prod_{j=k+1}^{n_1} C^{(1)}_{a_j}(\1l{j})\vac\ G^{a_{n_1}\ldots
a_1}\cr
&\qquad \times\ (-1)^{\ge_a\sum_{l=1}^{k-1}\ge_{a_l}}\cr
&= \sum_{k=1}^{n_1} \prod_{i=1}^{k-1} C^{(1)}_{b_i}(\1l{i})
\prod_{j=k+1}^{n_1} C^{(1)}_{b_{j}}(\1l{j})\vacn
\left({\Omega^{(1)}_k}\right)^{b_1\ldots b_{k-1}ab_{k+1}\ldots b_{n_1}}
(-1)^{\ge_a\sum_{l=1}^{k-1}\ge_{a_l}}
\quad .\cr}}

\equation{limS}{S = \sum_{k=1}^L e_k^{21} = i\
\lim_{\mu\rightarrow\infty} \mu\ A_{12}(\mu)}

\equation{limS2}{\eqalign{
S\ |\l{1}\ldots\l{n}|F\rangle &=
i\ \lim_{\mu\rightarrow\infty}\mu\ A_{12}(\mu)
|\l{1}\ldots\l{n}|F\rangle \cr
&= i\prod_{l=1}^nC_{b_l}(\l{l})\vac\
\lim_{\mu\rightarrow\infty}\mu\left({A^{(1)}_{12}(\mu)}\right)^{b_1\ldots
b_n}_{a_1\ldots a_n} \ F^{a_1\ldots a_n}\cr
&= i\prod_{l=1}^nC_{b_l}(\l{l})\vac\
\prod_{k=1}^{n_1}C^{(1)}_{b_k}(\1l{k})\vacn\
\lim_{\mu\rightarrow\infty}\mu\left({T^{(2)}_{n_1}(\mu)}^{12}
G\right)^{c_{n_1}\ldots c_1}
\quad .\cr}}

\equation{limS3}{
0 = i\ \lim_{\mu\rightarrow\infty}\mu
B^{(2)}(\mu)|\2l{1}\ldots\2l{n_2}\rangle =:\
J^{(2)}|\2l{1}\ldots\2l{n_2}\rangle\quad .}

\equation{limS4}{\left\lbrack J^{(2)}, C^{(2)}(\2l{})\right\rbrack
= D^{(2)}(\2l{})-A^{(2)}(\2l{})\quad ,}


\pagenumstyle{blank}
\footnoteskip=2pt
\line{\it September 1992\hfil ITP-SB-92-57}
\line{\it \hfil PUPT-1352}
\vskip4em
\baselineskip=32pt
\begin{center}{\bigsize{\sc Exact Solution of an electronic model
of\\ superconductivity in $1+1$ dimensions I}
{\baselineskip=20pt\footnote{This work was supported in part by
the National Science Foundation under research grants PHY91-07261 and
NSF91-08054.}}}
\end{center}
\vfil
\baselineskip=16pt

\begin{center}
{\bigsize
Fabian H.L.E\sharps ler\footnote[$\ \flat$]{\sc e-mail:
fabman@max.physics.sunysb.edu}\vskip .5cm
Vladimir E. Korepin\footnote[$\ \sharp$]{\sc e-mail:
korepin@max.physics.sunysb.edu}}
\vskip .5cm
{\it Institute for Theoretical Physics\vskip 4pt
State University of New York at Stony Brook\vskip 4pt
Stony Brook, NY~~11794-3840}
\vskip .3cm
and
\vskip .3cm
{\bigsize Kareljan Schoutens\footnote[$\ \natural$]{\sc e-mail:
schouten@puhep1.princeton.edu}}
\vskip .5cm
{\it Joseph Henry Laboratories\vskip 4pt
Princeton University\vskip 4pt
Jadwin Hall, P.O. Box 708\vskip 4pt
Princeton, NJ~~08544-0708}
\end{center}

\vfil

\centertext{\bfs \bigsize ABSTRACT}
\vskip\belowsectionskip

\begin{narrow}[4em]
We study a superconducting integrable model of strongly correlated
electrons in $1+1$ dimensions. We construct all six Bethe
Ans\"atze for the model and give explicit expressions for lowest
conservation laws. We also prove a lowest weight theorem for the
Bethe-Ansatz states.

\end{narrow}

\vfil

\break


\footnotenumstyle{arabic}
\sectionstyle{left}
\pagenumstyle{arabic}
\pagenum=0
{\bfs\section{Introduction}}

The Hubbard model has been studied as the prime example of a
microscopic theory of strongly correlated electrons for several
decades. More recently it was proposed as a model for
superconductivity by P.W. Anderson. It seems to be clear that in order
to make close contact to experimental results, the Hubbard hamiltonian
has to be modified by adding competing interactions. Despite the
simple form of its hamiltonian, the mathematical structure of the
Hubbard model is rather complex (as can be seen in the Bethe Ansatz
solution in one spatial dimension) and exact results about its
physical features are quite difficult to obtain. The ground states in
more than one dimension are still unknown, and the question under what
conditions the model exhibits superconductivity is yet unanswered.
Therefore it would be useful to have a model similar to the Hubbard
model, where the physics is more transparent.
In a recent publication we have presented a new model,
describing strongly correlated electrons on a general $d$-dimensional
lattice\upref eks/, which differs from the Hubbard model by moderate
interactions of nearest neighbours. The nature of these interactions
is very similar to the ones of Hirsch's model of hole
superconductivity \upref hirsch/.
In one dimension we the model is
integrable. We gave an exact expression for the ground state in the
attractive regime in $1$, $2$ and $3$ dimensions and showed that the
ground states of the model for attractive and moderately repulsive
on-site interaction exhibit off-diagonal long range order (ODLRO) and
are thus superconducting\upref eks2/.
In order to gain further insights into the physical features of the
model it is reasonable to first study its one dimensional version,
where integrability permits us to obtain exact results.
In this paper we construct the nested Algebraic Bethe Ansatz (NABA)
for the model and use it to determine higher conservation laws.
We also prove a lowest weight theorem for the Bethe Ansatz
states.\\
In two forthcoming publications we will determine the structure of the
low lying excitations over the ground states and
prove completeness of the $u(2|2)$ extended Bethe Ansatz\upref eks3/.

Electrons on a lattice are described by operators $c_{j,\sigma}\ $,
$j=1,\ldots,L$, $\sigma=\pm 1$, where $L$ is the total number
of lattice sites. These are canonical Fermi operators with
anti-commutation relations given by $\{ c^\dagger_{i,\sigma} ,
c_{j,\tau} \} = \delta_{i,j} \delta_{\sigma,\tau}$. The state
$\vac $ (the Fock vacuum) satisfies $c_{i,\sigma} \vac = 0$.
There are four possible electronic states at a given lattice site $i$
$$\putequation{config}$$
By $n_{i,\sigma}= c^\dagger_{i,\sigma} c_{i,\sigma}$ we denote the number
operator for electrons with spin $\sigma$ on site $i$ and we write
$n_i=n_{i,1} + n_{i,-1}$. \\
The hamiltonian on a lattice of $L$ sites in the grand canonical
ensemble is
$$\putequation{h}$$
where $H^0=-\sum_{j=1}^L H^0_{j,j+1}$ is given by
$$\putequation{hnaught}$$
Here $U$ is the Hubbard model coupling constant, $h$ is an external
magnetic field, and $\mu$ is the chemical potential.\\
It was shown in [\putref{eks}] that $H^0$ is invariant under a
$u(2|2)$ symmetry algebra. The three other terms in
$H$ are elements of the Cartan subalgebra of $u(2|2)$ (and commute
with $H^0$ and each other), and thus $H^0$ and $H$ have a complete set
of simultaneous eigenstates. In what follows we will construct a
complete set of eigenstates of $H^0$ (and of $H$) by using a
nested Algebraic Bethe Ansatz and the $u(2|2)$ symmetry of $H^0$.
The algebra $u(2|2)$ has a total of $16$ generators, $8$ of which are
fermionic and bosonic, respectively. The eight bosonic operators fall
into two $su(2)$ and two $u(1)$ subalgebras.
The two $su(2)$ algebras are generated
by the spin-operators [We shall always give local expressions
${\cal O}_j$ for symmetry generators, implying that the global ones
are obtained as ${\cal O} = \sum_{j=1}^L {\cal O}_j$.]
$$\putequation{spin}$$
and by the $\eta$-pairing like operators\upref yeta/
$$\putequation{eta}$$
Note that unlike for the case of the Hubbard model as treated in
[\putref{yeta}] there is no factor of $(-1)^j$ in the definition of
$\eta_j$.
The two $u(1)$ charges are given by the identity operator and the
operator
$$\putequation{X}$$
As all eigenvalues of $S^z+\eta^z = -{\hat N}_{-1}+{L\over 2}$ (where
${\hat N}_{-1} = \sum_{j=1}^L n_{j,-1}$ is the number operator for
spin down electrons) are either integer (for even $L$) or half-odd
integer (for odd $L$), the two $su(2)$ algebras are not quite
independent. Although all six generators of the two $su(2)$ algebras
commute with $H^0$, the Hilbert space of eigenstates of $H^0$ will
only carry representations of $su(2)\times su(2)$ with
integer (for even $L$) or half-odd integer (for odd $L$) quantum
numbers. The same kind of restriction carries through to the complete
$u(2|2)$ symmetry.
The eight fermionic generators are given by
$$\putequation{ferm}$$
and their hermitean conjugates.

We have shown in [\putref{eks}] that $H^0$ can be written as minus the
sum over graded permutation operators
$$\putequation{hpi}$$
where the operator $\Pg ^{j,j+1}$ permutes
the four possible configurations (\putlab{config})
between the sites $j$ and $j+1$, picking up a minus sign if both of
the permuted configurations are fermionic, {\sl i.e.}
$$\putequation{perm}$$
It is clear that this form of interaction conserves the individual
numbers $N_\uparrow$ and $N_\downarrow$ of electrons with spin up
and spin down, and the numbers $N_l$ and $N_h$ of doubly occupied
(``local electron pairs'') and empty sites (``holes'').
We will choose the following conventions throughout this paper
\item {} $N_\up$ = number of single electrons with spin up
\item {} $N_\da$ = number of single electrons with spin down
\item {} $N_e=N_\up+N_{\da}$ = number of single electrons
\item {} $N_l$ = number of local electron pairs
\item {} $N_h$ = number of holes
\item {} $N_b=N_h+N_l$ = number of ``bosons''.
\vskip .5cm

The outline of this paper is as follows :\hfill\break
In section $2$ we perform a detailed construction of the
Algebraic Bethe Ansatz of the model.
We derive six different forms for the Bethe Ansatz Equations (BAE)
and the eigenvalues of the transfer matrix.
The graded Quantum Inverse Scattering Method (QISM), discussed in
section $2$, enables us to obtain expressions for the hamiltonian and
(an infinite number of) higher conservation laws at the quantum level.
These conserved charges are of interest, because physical
interactions, although of short range, are not generally well
approximated by interactions involving only nearest neighbours.
The charges under consideration involve interactions of longer range
(next nearest neighbours, next next nearest {\sl etc.})
and can be added to the hamiltonian without spoiling the
integrability of the model. Thus it is possible to construct
integrable models with longer range interactions by using higher
conservation laws\upref Frahm/.
The first nontrivial higher integral of motion is for example given by
the expression
$$\putequation{hintro}$$
where $H^0_{k,k+1}$ is the density of the hamiltonian defined in
(\putlab{h}).
Section $3$ is devoted to the derivation of explicit formulas for
higher conservation laws.

{\bfs\section{Graded Quantum Inverse Scattering Method}}
In this section we discuss how to embed the hamiltonian
(\putlab{hnaught}) into the framework of the Quantum Inverse
Scattering Method (QISM). For supersymmetric theories it is necessary
to modify the QISM along the lines discussed by Kulish and Sklyanin in
[\putref{Kul, SklKul}].
In [\putref{ek}] we gave a summary of the ``graded'' version of the
QISM, so that we will constrain ourselves to a very brief discussion
below and refer to [\putref{ek}] for more details.
Starting point of the QISM is an $R$-matrix, which obeys a graded Yang
Baxter equation. From this $R$-matrix we then construct a family of
commuting transfer matrices, describing a ``fundamental'' spin model.
The transfer matrix is the generating functional of an infinite number
of conservation laws, the second of which we show to be the
hamiltonian (\putlab{hnaught}).
Finally we construct a set of simultaneous eigenstates of the
transfer matrix and the hamiltonian, using a nested Algebraic Bethe
Ansatz (NABA)\upref Kul,KulRe,KulRe2,Taktajan,ek/.
Due to the grading there exist {\sl a priori} $4!=24$ choices for the
$R$-matrix, all of them describing the same physical system, but
leading to different (yet equivalent as shown in Appendix B) forms of
the NABA. As the model has the two trivial discrete symmetries of
``spin reflection'' (interchange of spin up and spin down electrons)
and of interchange of empty sites and doubly occupied sites,
there are essentially only six different solutions\footnote{Each of
these solutions generates a ``multiplet'' of four NABA {\sl via} the
substitutions $N_\up\leftrightarrow N_\down$ and
$N_l\leftrightarrow N_h$}. One of the six solutions has been
previously obtained in [\putref{Kul}].
By the same kind of reasoning one can show that there are
${(n+m)!\over n!\ m!}$ Bethe Ansatz solutions for a permutation-type
model with $u(m|n)$ symmetry.

\subsection{2.1.\ \sl Yang Baxter Equation}

Consider the graded linear space $V^{(m|n)}$, where
$m$ and $n$ denote the dimensions of the ``even'' and ``odd''
parts respectively.
A basis $\lbrace e_1,\ldots ,e_{m+n}\rbrace$ of $V^{(m+n)}$
consists of $m+n$ vectors $e_a$ with Grassmann parities $\ge_a$ such
that for $m$ vectors $\ge_a=0$ and for $n$ vectors $\ge_a=1$.
The grading of the matrix elements $M_{ab}$ of linear operators
$M$ on $\vmn$ (in the basis $\lbrace e_1,\ldots ,e_{m+n}\rbrace$) is
given by $\ge_{M_{ab}} = \ge_a+\ge_b$.
The supertrace of $M$ is defined as
$$\putequation{str}$$

The graded tensor product space $V^{(m|n)}\otimes V^{(m|n)}$ is given
in terms of its basis vectors $\lbrace e_a\otimes e_b|a,b=1,\ldots,
m+n\rbrace$ as follows
$$\putequation{tensorprod}$$
The action of the linear operator $F\otimes G$ on the vector
$v\otimes w$ in $\vmn\otimes\vmn$ is defined by
$$\putequation{linop}$$
which results in matrix elements of the form
$$\putequation{linopmat}$$
The identity operator $I$ in $\vmn\otimes\vmn$ and the matrix $\Pg$
that permutes the individual linear spaces in the tensor product space
are given by

$$\putequation{pigr}$$

A matrix $R(\l{})$ (depending on a spectral parameter $\l{}$) is said
to fulfill a graded Yang-Baxter-equation, if the
following identity on $\vmn\otimes \vmn\otimes \vmn$ holds
$$\putequation{YBE1}$$
In components this identity reads
$$\putequation{YBE}$$
Note that despite the fact that the tensor product in (\putlab{YBE1})
carries a grading, there are no extra signs in (\putlab{YBE}) compared
to the nongraded case.
It is easily checked that the R-matrix
$$\putequation{Rgen}$$
fulfills equation (\putlab{YBE}).

\subsection{2.2.\ \sl Construction of the Transfer Matrix}
\subsectionnum=2

{}From (\putlab{YBE}) one can derive the equation
$$\putequation{YBE2}$$
where the indices $1,2,3$ indicate in which of the spaces $\vmn$ in
the tensor product space $\vmn\otimes \vmn\otimes \vmn$ the matrices
act nontrivially.
The tensor product in (\putlab{YBE2}) is between the spaces $1$ and
$2$.  We now call the third space ``quantum space''and the first two
spaces ``matrix spaces''.
The physical interpretation of the quantum space is as the Hilbert
space over a single site of a one-dimensional lattice. We now consider
the situation, where intertwining relations of the type
(\putlab{YBE2}) hold for all sites of a lattice of length $L$. The
quantum space index ``$3$'' now gets replaced by an index labelling
the number of the site.
We define the $L$-operator (on site $n$) as a linear operator on
${\cal H}_n\otimes V^{(m|n)}_{matrix}$ (where ${\cal H}_n\simeq \vmn$
is the Hilbert space over the $n^{th}$ site, and $V^{(m|n)}_{matrix}$ is
a matrix space)
$$\putequation{LR}$$
$L_n$ is a quantum operator valued $(m+n)\times (m+n)$ matrix, with
quantum operators acting nontrivially in the $n^{th}$ quantum space
(of the direct product Hilbert space over the complete lattice
$\otimes_{j=1}^L {\cal H}_j$).
The greek indices are the ``quantum indices'' and the roman indices
are the ``matrix indices''.
Equation (\putlab{YBE2}) for the $n^{th}$ quantum space can now be
rewritten as the operator equation
$$\putequation{int2}$$
Here the graded tensor product is between the two matrix spaces and
$R$ only acts in the matrix spaces.
The intertwining relation (\putlab{int2}) enables us to construct an
integrable spin model as follows.\hfill\break
We first define the monodromy matrix $T_L(\l{})$ as the matrix product
over the $L$-operators on all sites of the lattice, {\sl i.e.}
$$\putequation{mon}$$
$T_L(\l{})$ is a quantum operator valued $(m+n)\times (m+n)$ matrix
that acts nontrivially in the graded tensor product of all quantum
spaces of the lattice and by construction fullfills the same
intertwining relation as the $L$-operators (as can be proven by
induction over the length of the lattice)
$$\putequation{intT}$$
The transfer matrix $\tau (\l{})$ of the integrable spin model is now
given as the matrix supertrace of the mondromy matrix
$$\putequation{tau}$$
As a consequence of (\putlab{intT}) transfer matrices with different
spectral parameters commute. This condition implies that
the transfer matrix is the generating functional of
the hamiltonian and of an infinite number of ``higher'' conservation
laws of the model.

\subsection{2.3.\ \sl Trace Identities}
\subsectionnum=3

Taking logarithmic derivatives of the transfer matrix at a special
value of the spectral parameter, one can generate higher conservation
laws\upref{Vlad}/.
For the transfer matrix constructed from the $R$-matrix
(\putlab{Rgen}) along the lines discussed in section $2.2.$, the
corresponding hamiltonian is obtained by taking the first logarithmic
derivative at zero spectral parameter
$$\putequation{trid1}$$
The proof of this identity can be carried out in the same way as for
the ungraded case, the main difference being the grading
of the tensor product of the quantum spaces (see (\putlab{mon})).
By shifting the energy eigenvalues by a constant we
obtain the expression (\putlab{hpi}) for the hamiltonian
(\putlab{hnaught})
$$\putequation{trid2}$$
if we choose our underlying graded vector space to have signature
$(2,2)$, {\sl i.e.} to have a basis with two fermionic and two bosonic
states.
This shows that the transfer matrix constructed from the $L$-operator
(\putlab{LR}) and $R$-matrix (\putlab{Rgen}) is indeed the correct
transfer matrix for the hamiltonian (\putlab{hnaught}).
Higher conservation laws are obtained as the coefficients of the power
series
$$\putequation{hcl}$$
There exists however a simpler method for the construction of higher
integrals of motion than taking logarithmic derivatives, which we will
discuss in section $3$.
For fundamental spin models (and thus for the model at hand) the
transfer matrix at the special value of the spectral parameter (which
is zero for our case) is equal to the translation operator. Therefore
we can construct the momentum operator from the logarithm of the
transfer matrix at zero spectral parameter
$$\putequation{mtm}$$
By construction the momentum operator commutes with the hamiltonian
$H^0$ and all higher conservation laws $H_{(k+1)}$, {\sl i.e.}
$[P,H^0]=0=[P,H_{(k+1)}]$.
\subsection{2.4.\ \sl Algebraic Bethe Ansatz with a bosonic background
(FFBB grading)}

Due to the fact that there are four different configurations per site
for the model defined by (\putlab{h}), the Hilbert space at the
$k^{th}$ site of the lattice is isomorphic to ${\Complex}^4$ and is
spanned by the four vectors $e_1 = \pmatrix{1&0&0&0\cr}^T$, $e_2 =
\pmatrix{0&1&0&0\cr}^T$, $e_3 = \pmatrix{0&0&1&0\cr}^T$, and $e_4 =
\pmatrix{0&0&0&1\cr}^T$.
In this section we consider a grading such that
$e_1$ and $e_2$ are fermionic (representing spin down/spin up
electrons respectively) and $e_3$ and $e_4$ are bosonic (doubly
occupied/empty site). In terms of the Grassmann parities this means
that $\ge_1=\ge_2=1$ and $\ge_3 = \ge_4 = 0$.
We pick the reference state in the $k^{th}$ quantum space $\vac _k$,
and the vacuum $\vac$ of the complete lattice of $L$ sites, to
be the bosonic completely unoccupied state, {\sl i.e.}
$$\putequation{vacuum}$$
This choice of grading implies that the $L$-operator \pl{LR} is given by
the following expression
$$\putequation{Lop}$$
where $e_n^{ab}$ are quantum operators in the $n^{th}$ quantum space
with matrix representation $(e_n^{ab})_{\ga\gb} = \gd _{a\ga}\gd
_{b\gb}$.
The monodromy matrix (\putlab{mon}) is a quantum operator valued
$4\times 4$ matrix, which we represent as
$$\putequation{T}$$
The transfer matrix is then given by
$$\putequation{tau2}$$
The action of $L_k(\l{})$ on the reference state on the $k^{th}$ site
can be easily determined by using the matrix representation of the
$e_k^{ab}$ in \pl{Lop} and the vectorial representation of $\vac _k$
\pl{vacuum}
$$\putequation{Lvac}$$
Using (\putlab{mon}) and (\putlab{Lvac}) we determine the action of
the monodromy matrix on the reference state to be
$$\putequation{Tvac}$$
We will now construct a set of eigenstates of the transfer matrix
using the technique of the NABA.
Inspection of (\putlab{Tvac})
reveals that $C_i(\la )$ are creation operators with respect to our
choice of reference state.
This observation leads us to the following Ansatz for the eigenstates
of $\tau(\mu )$
$$\putequation{state1}$$
where the indices $a_j$ run over the values $1,2,3$, and
$F^{a_n\ldots a_1}$ is a function of the spectral parameters $\l{j}$.
The action of the transfer matrix on states of the form
(\putlab{state1}) is determined by (\putlab{Tvac}) and the
intertwining relations (\putlab{intT}).
The components of the intertwining relations (\putlab{intT}) needed
for the construction of the NABA are
$$\putequation{AC}$$
where
$$\putequation{r}$$
Here ${\P^{(1)}}^{ab}_{cd}$ is the $9\times 9$ permutation matrix
corresponding to the grading $\ge_1=\ge_2=1\ ,\ \ge_3=0$.
$r(\mu)$ can be seen to fulfill a (graded) Yang-Baxter equation on its
own
$$\putequation{ybe}$$
and can be identified with the $R$-matrix of a fundamental spin model
describing two species of fermions and one species of bosons.
Using (\putlab{AC}) we find that
the diagonal elements of the monodromy matrix act on the states
(\putlab{state1}) as follows
$$\putequation{Dstate}$$
$$\putequation{Astate}$$
where
$$\putequation{tau1}$$
Note that the tensor products between the $n$ quantum spaces in the
expression for the operator $\tau^{(1)}(\mu)$ are once again graded,
which results in minus signs given by \pl{mon}.
$L^{(1)}$ and $r(\mu)$ can be
interpreted as $L$-operator and $R$-matrix of a fundamental spin model
($r$ fulfills the Yang-Baxter equation (\putlab{ybe})), describing two
species of fermions and one species of bosons. Hence $T_n^{(1)}(\mu )$
and $\tau^{(1)}(\mu )$ are the monodromy matrix and transfer matrix of
the corresponding {\sl inhomogeneous} model.
This model can be identified with the inhomogeneous supersymmetric
$t$-$J$ model on a lattice with $n$ sites. The Algebraic
Bethe Ans\"atze for the supersymmetric $t$-$J$ model were constructed
in detail in [\putref{ek}], so that we will restrict ourselves to a
abbreviated discussion and refer to [\putref{ek}] for derivations of
some of the results. Inspection of (\putlab{Dstate}) and
(\putlab{Astate}) together with (\putlab{tau2}) shows that the
eigenvalue condition
$$\putequation{ew}$$
leads to the requirements that $F$ ought to be an eigenvector of the
``nested'' transfer matrix $\tau^{(1)}(\mu )$, and that the ``unwanted
terms'' cancel, {\sl i.e.}
$$\putequation{unw}$$
The notation $\nu(\mu)$ for the eigenvalues of the transfer matrix
$\tau(\mu)$ is to be considered as a shorthand as the eigenvalues
depend also on the spectral parameters $\l{j}$ and coefficients $F$.
The quantities $\Lambda_k$ and $\tilde\Lambda_k$ are computed in
Appendix A. Using their explicit expressions in (\putlab{unw}) we
obtain the following conditions on the spectral parameters
$\l{j}$ and coefficients $F$, which are necessary for (\putlab{ew}) to
hold
$$\putequation{PBC}$$
This completes the first step of the NABA. In the next step we will
now solve the first nesting.
The condition that $F$ ought to be an eigenvector of $\tau ^{(1)}(\mu )$
requires the diagonalisation of $\tau ^{(1)}(\mu )$, which can be carried
out by a second, ``nested'' Bethe Ansatz.
{}From (\putlab{ybe}) and (\putlab{tau1}) the following intertwining
relation is easily derived
$$\putequation{intnest}$$
If we write
$$\putequation{T1}$$
then (\putlab{intnest}) and (\putlab{r}) imply that
$$\putequation{intnest2}$$
Here the roman indices take the values $1,2$, both of which are
fermionic ($\ge_1 = 1 = \ge_2$).
The $R$-matrix $r^{(1)}(\mu)$ can again be proven to fulfill a
Yang-Baxter equation of the form \pl{ybe}.
As the reference state for the first nesting we pick
$$\putequation{vacnest}$$
The action of the nested monodromy matrix $T^{(1)}_n(\mu )$ on the
reference state $\vacn $ is determined by (\putlab{tau1}) and we find
$$\putequation{tvac}$$
We now make the following Ansatz for the eigenstates of
$\tau^{(1)}(\mu )$
$$\putequation{state2}$$
Here the indides $b_i$ can take the two values $1$ and $2$.
These states can be related to the coefficients $F^{a_n\ldots
a_1}$ in the following way :
The state ${|\1l{1},\ldots , \1l{n_1}\rangle}$ ``lives'' on a lattice
of $n$ sites and is thus an element of a direct product over $n$
Hilbert spaces. In components it reads ${|\1l{1}\ldots
\1l{n_1}\rangle}_{a_n\ldots a_1}$, which can be directly identified
with $F^{a_n\ldots a_1}$.\hfill\break
The action of $\tau^{(1)}(\mu )$ on the states (\putlab{state2}) can
be evaluated with the help of (\putlab{intnest2}) and \pl{tvac}
$$\putequation{Dstate1}$$
$$\putequation{Astate1}$$
Here $\tau^{(2)}(\mu)$ is the transfer matrix of the second nesting
$$\putequation{tau3}$$
The operator $\tau^{(2)}(\mu)$ can be interpreted as the transfer
matrix of an inhomogeneous model on $n_1$ sites describing two species
of fermions.
It is easily seen from \pl{Astate1} that for the states \pl{state2} to
be eigenstates of the $\tau^{(1)}(\mu)$, $G$ must be an eigenvector of
$\tau^{(2)}(\mu)$ with eigenvalue $\nu^{(2)}(\mu)$.
{}From (\putlab{Dstate1}) and (\putlab{Astate1})
one can read off the eigenvalues of
$\tau^{(1)}(\mu )$
$$\putequation{eig}$$

The unwanted terms $\Lambda^{(1)}_k$ and $\tilde\Lambda^{(1)}_k$ are
computed in Appendix A and their cancellation (which ensures that the
states (\putlab{state2}) are eigenstates of the transfer matrix
$\tau^{(1)}(\mu )$ ) leads to the following set of Bethe Ansatz
equations (BAE) for the first nesting
$$\putequation{pbc1}$$
The diagonalisation of $\tau^{(2)}(\mu )$ can be performed by a third
Bethe Ansatz. From the Yang-Baxter relation for $r^{(2)}(\mu)$
one can derive the following intertwining relation
$$\putequation{intnest3}$$
The components of \pl{intnest3} needed for the construction of an
Algebraic Bethe Ansatz are
$$\putequation{intnest4}$$
As the reference state we pick
$$\putequation{vacnest2}$$
The action of the nested monodromy matrix $T^{(2)}_{n_1}(\mu )$ on the
reference state $\vacnn $ is determined by (\putlab{tau3}) and we find
$$\putequation{tvac2}$$
We now make the following Ansatz for the eigenstates of
$\tau^{(2)}(\mu )$
$$\putequation{state3}$$
The analysis now proceeds analogous to the one for the first nesting.
The eigenvalues of $\tau^{(2)}(\mu)$ on the states \pl{state3} are
found to be
$$\putequation{eig2}$$
under the condition that the spectral parameters $\2l{p}$ are solutions
to the BAE
$$\putequation{pbc2}$$
Using \pl{eig2} in \pl{pbc1} ($G$ is an eigenvector of
$\tau^{(2)}(\1l{k})$ with eigenvalue $\nu^{(2)}(\1l{k})$ ) we obtain
the BAE for the first nesting. If we then insert \pl{eig2} in \pl{eig}
and use the resulting expression for the eigenvalues of
$\tau^{(1)}(\mu)$ in \pl{PBC}, we obtain the complete set of three
BAE. If we shift the spectral parameters according to
$$\putequation{tildeFFBB}$$
these equations read
$$\putequation{BAEFFBB}$$
Here we have used that $n=N_e+N_l=N_\da+N_\up+N_l$, $n_1 = N_e$, $n_2
= N_\da$, which follow from our identification of the four $\Complex
^4$ basis vectors with the four possible configurations per site
\pl{config}.
The eigenvalues of the transfer martix $\tau(\mu)$ are computed by
using only the wanted terms in \pl{Astate} and \pl{Dstate}, and
inserting \pl{eig} and \pl{eig2} in the resulting expression
$$\putequation{nuFFBB}$$
Note that the condition of cancellation of the residues of the poles
in \pl{nuFFBB} implies the BAE \pl{BAEFFBB}.
The states \pl{state1} are simultaneous eigenstates of $H^0$ and
$\tau (\mu)$, and the energy eigenvalues can be determined from the
eigenvalues of the transfer matrix \pl{nuFFBB} by using the trace
identity \pl{trid2}.
We find
$$\putequation{Effbb}$$
where we have reparametrised $\tl{j}={1\over 2}cot({k_j\over 2})$.
The eigenvalues of the momentum operator (\putlab{mtm}) can be
obtained by setting $\mu =0$ in the expression (\putlab{nuFFBB}) for
the eigenvalues of the transfer matrix, and then taking the logarithm.
We find
$$\putequation{pFFBB}$$

\subsection{2.5.\ \sl Algebraic Bethe Ansatz for the BBFF grading}
The construction of the NABA for this choice of grading follows
roughly the same lines as for the FFBB case. Due to the fact that we
are ``truncating'' fermionic lines, there are additional
complications on the first two steps of the NABA. The situation is
analogous to the one for the BFF grading in the supersymmetric
$t$-$J$ model, which was discussed in [\putref{ek}]. The necessary
steps are all explained in that paper, so that we will constrain
ourselves to only a brief sketch of the derivations, and mainly just
quote the results.

The grading is chosen in such a way that
$e_3$ and $e_4$ are fermionic (representing spin down/spin up
electrons, respectively) and $e_1$ and $e_2$ are bosonic (doubly
occupied/empty site). In terms of the Grassmann parities this means
that $\ge_3=\ge_4=1$ and $\ge_1 = \ge_2 = 0$.
The reference state in the $k^{th}$ quantum space $\vac _k$
and the vacuum $\vac$ of the complete lattice of $L$ sites are
fermionic with all spins up, {\sl i.e.}
$$\putequation{vacuum2}$$
The choice of grading leads to the L-operator
$$\putequation{Lop2}$$
We again partition the monodromy matrix like in \pl{T}, which implies
the following expression for the transfer matrix
$$\putequation{tau22}$$
The action of the monodromy matrix on the vacuum can be computed using
\pl{vacuum2} and \pl{Lop2}
$$\putequation{Tvac2}$$
As before we identify $C_a$ as creation operators with respect to our
choice of reference state and make an Ansatz of the type \pl{state1}
for the eigenstates of the transfer matrix \pl{tau22}.
The relevant intertwining relations are
$$\putequation{AC2}$$
where
$$\putequation{r2}$$
Here $\Pi_{FFB}$ and $\Pi_{BBF}$ are the permutation matrices for the
gradings $\ge_1=\ge_2=1\ ,\ \ge_3=0$ and $\ge_1=\ge_2=0\ ,\ \ge_3=1$
respectively.
The intertwining relations (\putlab{AC2}) imply the following action
of the diagonal elements of the monodromy matrix on the states
$|\l{1},\ldots , \l{n}|F\rangle = C_{a_1}(\l{1})\ C_{a_2}(\l{2})\ldots
C_{a_n}(\l{n})\ \vac\  F^{a_n\ldots a_1}$
$$\putequation{Dstate2}$$
$$\putequation{Astate2}$$
where
$$\putequation{tausuth}$$
Here all indices $c_i$ and $c$ are summed over.
The $L$-operators $L^{(1)}(\la)$ are constructed by multiplying the
R-matrix $r_{BBF}(\la)$ by the permutation matrix $\Pi_{BBF}$ and are found
to be
$$\putequation{lnest21}$$
The expression for $\tau^{(1)}(\mu )$ is significantly different from
the one in the $FFBB$ case, but by using an appropriate definition for
the graded tensor product $\tau^{(1)}(\mu)$ can again be interpreted
as the transfer matrix of an inhomogeneous spin model on a lattice of
$n$ sites. Our reference state $\vac$ is now of fermionic nature and
we have to define a new graded tensor product reflecting this fact
$$\putequation{newgrtp}$$
Effectively the new graded tensor product switches even and odd
Grassmann parities, {\sl i.e.} $\ge_a\longrightarrow \ge_a +1$.
In terms of this tensor product the transfer matrix $\tau^{(1)}(\mu )$
given by (\putlab{tausuth}) can be rewritten as
$$\putequation{tau12}$$
In the expression after the second equality of (\putlab{tau12}) we
have explicitly written the tensor product $\osl$ between the quantum
spaces over the sites of the inhomogeneous model (the $L$-operators
are of course again multiplied as matrices).
As before $F^{a_n\ldots a_1}$ must be an eigenvector of
$\tau^{(1)}(\mu )$, if $|\l{1}\ldots \l{n}|F\rangle$ is to be an
eigenstate of $\tau (\mu)$.
The unwanted terms are computed in Appendix B, and the
condition of their cancellation
$$\putequation{cancelbff}$$
leads to the conditions
$$\putequation{pbcbff1}$$

In the solution of the first nesting there is an additional
complication compared to the FFBB case.
Because of our change of tensor product, the $L$-operators
$L^{(1)}(\la)$ are not intertwined by the R-matrix $r(\mu)$ defined in
(\putlab{r2}), but by the R-matrix
$$\putequation{otherr}$$
The following intertwining relation holds
$$\putequation{intnestbff1}$$
If we make the following choice of reference state for the first
nesting
$$\putequation{vacnest22}$$
one can show that the model defined by \pl{intnestbff1},
\pl{vacnest22} and the graded tensor product $\osl$ is isomorpic to a
model of the permutation type with $BBF$ grading (describing two
species of bosons and one species of fermions) and the graded tensor
product $\otimes$. Effectively the first nesting thus describes the
inhomogeneous model obtained by just ``truncating a fermionic line''
from the original model. This holds generally for $u(m|n)$ symmetric
models\upref fab/.\par
In the next step of the NABA we thus face the task of solving the
inhomogeneous BBF model. As in the first step we have to truncate a
fermionic line, which can be done in a way analogous to the procedure
described above. The intertwining relations are of the same form as
\pl{AC2}, but the R-matrices are now the ones of the BB and FF
models respectively. Taking only the wanted terms into account we
obtain the eigenvalues of the transfer matrix $\tau^{(1)}(\mu)$ as
given below. The condition of the cancellation of the unwanted terms
now reads
$$\putequation{pbcbff2}$$
where we have made an Ansatz of the form \pl{state2} for the
eigenstates of the transfer matrix $\tau^{(1)}(\mu)$, and where
$\tau^{(2)}(\mu)$ is the transfer matrix of the second nesting given
by the following expression
$$\putequation{tausuth2}$$
Here $L^{(2)}(\mu) = a(\mu) I^{(2)} + b(\mu) \Pi_{BB}$.
Note that the roman indices in \pl{pbcbff2} and \pl{tausuth2} can take
the values $1,2$, both of which carry bosonic grading and thus do not
generate any minus signs.
The second nesting can be shown to be equivalent to the
inhomogeneous spin ${1\over 2}$ Heisenberg XXX model, or BB model in
our terminology. The eigenvalues of $\tau^{(2)}(\mu)$ and the BAE for
the inhomogeneous BB model are easily determined, and we can put
together the three steps of the NABA like in the case of the FFBB
grading.
After reparametrising according to
$$\putequation{tildeBBFF}$$
the BAE take the form
$$\putequation{BAEBBFF}$$
The eigenvalues of the transfer matrix are given by
$$\putequation{nuBBFF}$$
Energy and momentum eigenvalues are given by
$$\putequation{epBBFF}$$
where we have defined $\tl{j} = {1\over 2} tan({k_j\over 2})$.
Note that after the reparametrisations \pl{tildeBBFF} and
\pl{tildeFFBB} the expressions for the BAE for the BBFF and FFBB
gradings coincide up to the substitutions $N_\da\leftrightarrow N_l$
and $N_\up\leftrightarrow N_h$. However there is no such obvious relation
between the eigenvalues of the transfer matrix.
For the energy eigenvalues it can be easily shown by taking
logarithmic derivatives of the eigenvalues $\nu(\mu)$, that the
expressions for BBFF and FFBB are identical up to the transformation
$N_\da\leftrightarrow N_l$ and $N_\up\leftrightarrow N_h$ apart from
an overall minus sign. This is clear as under a particle-hole
transformation for spin-up $H^0$ transforms into $-H^0$.
There is no such relation for the eigenvalues
of the momentum operator.\par
The BAE \pl{BAEBBFF} are particularly useful, because they reduce to
the BAE for the Sutherland solution of the supersymmetric $t$-$J$
model in the sector without localons ($N_l = 0$). The classification
of ground states and low lying excited states is particularly easy in
this case. In our forthcoming publication we will analyse the bound
state and ground state structure of this set of BAE, and describe the
model by a set of integral equations in the thermodynamic limit\upref
ek2/.

\subsection{2.6.\ \sl Algebraic Bethe Ansatz for the FBFB grading}
Here the first step of the NABA is completely analogous to the one for
the FFBB grading. The R-matrix and monodromy matrix of the nesting can
be identified with the ones for FBF grading of the inhomogeneous
supersymmetric $t$-$J$ model, which was solved in [\putref{ek}].
The eigenvalues of the transfer matrix are found to be
$$\putequation{nuFBFB}$$
Using the reparametrisation
$$\putequation{tildeFBFB}$$
the following form of the BAE is derived
$$\putequation{BAEFBFB}$$

\subsection{2.7.\ \sl Algebraic Bethe Ansatz for the BFBF grading}
In this choice of grading we first truncate a fermionic line
analogously to the first step of the BBFF case. In the first
nesting we truncate a bosonic line, which is done like in the first
step of the FFBB case. The resulting expression for the eigenvalues of
the transfer matrix are
$$\putequation{nuBFBF}$$
After reparametrising the spectral parameters according to
$$\putequation{tildeBFBF}$$
the BAE take the form
$$\putequation{BAEBFBF}$$
The BAE \pl{BAEBFBF} can be mapped onto the BAE \pl{BAEFBFB} by a
boson-fermion interchange like in the FFBB/BBFF case.

\subsection{2.8.\ \sl Algebraic Bethe Ansatz for the BFFB grading}
The results for the eigenvalues of the transfer matrix and BAE are
$$\putequation{nuBFFB}$$
$$\putequation{BAEBFFB}$$
where the we have reparametrised
$$\putequation{tildeBFFB}$$

\subsection{2.9.\ \sl Algebraic Bethe Ansatz for the FBBF grading}
This last possible choice of grading leads to the following results
$$\putequation{nuFBBF}$$
$$\putequation{BAEFBBF}$$
where
$$\putequation{tildeFBBF}$$
Again there exsists a boson-fermion interchange map between
\pl{BAEBFFB} and \pl{BAEFBBF}.

{\bfs{\section{Lowest Weight properties of the Bethe Ansatz states}}}
In this section we prove that all Bethe Ansatz states with finite
spectral parameters are lowest weight states of the global $u(2|2)$
symmetry algebra. For the case of the FFBB grading the lowest-weight
conditions read
$$\putequation{lws}$$
Lowest weight theorems hold for all choices of the grading (the
grading determines which generators annihilate the BA states),
but without loss of generality we will perform the computations only
for the specific case of the FFBB grading.
The lowest weight property is of great importance as it first of all
demonstrates that the Bethe Ansatz does not provide a complete set of
eigenstates of the hamiltonian $H^0$, and because it can be used
together with the global symmetry structure to generate a complete set
of eigenstates. Similar computations have been performed for other
models in [\putref{hubb,ft,kdv,fk}].

We start by noting that in the basis of the Hilbert space introduced
at the beginning of section $2.4$, all lowering generators of the
$u(2|2)$ algebra are represented in the form $\sum_{k=1}^L e_k^{ab}$.
By using the correspondence of the four basis vectors $e_i$ over one
site of the lattice with the four possible configurations \pl{config},
we find
$$\putequation{lowop}$$
where $e_k^{ij}$ are quantum operators acting nontrivially only on the
$k^{th}$ site of the lattice with matrix representation
$$\putequation{esubk}$$

Using the representation \pl{lowop} and the expression \pl{Lop} for
the $L$-operator we are able to rewrite graded quantum commutators as
matrix commutators
$$\putequation{lgen}$$
where $e^{ij}$ is a matrix of the form
$$\putequation{e}$$
Note that on the r.h.s of \pl{lgen} $e^{ij}$ and $L_k(\mu)$ are
multiplied as matrices.
The graded quantum commutator of $e_k^{ij}$ with
$\left(T_L(\mu)\right)^{ab}$ can now be expressed as a matrix
commutator by using \pl{lgen} and \pl{T}
$$\putequation{tgen}$$
Summing over all sites of the lattice we arrive at
$$\putequation{tgen2}$$
Equation \pl{tgen2} enables us to compute the graded commutators of
the $u(2|2)$-lowering operators with the creation operators
$C_a(\la)$. We find
$$\putequation{Cgen}$$
We also note that all lowering operators annihilate the vacuum
$$\putequation{annihilate}$$

{\sl \subsection{Lowest weight property for $Q_1$, $Q_{-1}$ and
$\eta$}}
In this subsection we show that all Bethe Ansatz eigenstates are
are annihilated by the $u(2|2)$ generators $J_1 = Q_{-1}$, $J_2 =
Q_{1}$, and $J_3 = \eta$.
Using \pl{annihilate} and \pl{Cgen} we find that
$$\putequation{lw1}$$
The last equality follows by inspection of the r.h.s. of the previous
line, as $D(\l{k})$ and $A_{aa_k}(\l{k})$ can be moved past
the string of $C_b$'s by using \pl{AC}.
Due to the cancellation of the unwanted terms (which are explicitly
computed in Appendix A) all of the quantities $\Omega_k$ vanish as we
will demonstrate below.\\
{}From \pl{lw1} it follows that $\Omega_1$ is the coefficient of the
term, where the creation operator $C_b$ with spectral parameter
$\l{1}$ is missing. It can be obtained from the term
$$\putequation{lw2}$$
by moving $(\gd_{aa_1}D(\l{1})-A_{aa_1}(\l{1}))$ past
$\prod_{j=2}^n C_{a_j}(\l{j})$ by only taking ``wanted'' terms into
account, {\sl i.e.,} by only using the first contribution of the
intertwining relations \pl{AC} when moving $A_{ab}(\l{1})$ or
$D(\l{1})$ past $C_{b_j}(\l{j})$. The resulting expression for
$\Omega_1$ is found to be
$$\putequation{o1}$$
If we compare \pl{o1} with the condition of the cancellation of the
unwanted terms \pl{laktilde} and \pl{lak1}, we find that $\Omega_1$
vanishes because
$$({\tilde\Lambda}_1F)^{ab_2\ldots b_n}+(\Lambda_1F)^{ab_2\ldots b_n}
=0\quad .$$
The expression for $\Omega_k$ can be obtained in a similar fashion
:\hfill\break
We use \pl{AC} to rewrite $|\l{1}\ldots \l{n}\rangle$ as
$$\putequation{kfront}$$
Now the action of the lowering operators $J_a$ on the Bethe Ansatz
states can be written in the form
$$\putequation{lw3}$$
Going through the same steps as in \pl{lw1} it follows that $\Omega_k$
can be obtained in essentially the same way as $\Omega_1$ with the
result
$$\putequation{ok}$$
Inspection of the \pl{lak2} and \pl{laktilde} in Appendix A shows that
the condition of the cancellation of the unwanted terms
$$({\tilde\Lambda}_kF)^{b_1\ldots b_{k-1}ab_{k+1}\ldots
b_n}+(\Lambda_kF)^{b_1\ldots b_{k-1}ab_{k+1}\ldots b_n} =0$$
implies the vanishing of $\Omega_k$. This completes the proof of the
lowest weight property of the Bethe Ansatz states for the generators
$Q_1$, $Q_{-1}$ and $\eta$.
\vskip .5cm

{\sl \subsection{Lowest weight property for $\tqd_1$ and $\tqd_{-1}$}}
In this subsection we show that the lowering generators $\tqd_{\pm 1}$
annihilate all Bethe Ansatz states with finite spectral parameters
$\l{j}$. We first reexpress $\tqd_{\pm 1}$ in terms of matrix elements
of the monodromy matrix \pl{T}
$$\putequation{lim}$$
Using the intertwining relation \pl{AC} we are
able to determine the action of $\mu\ A_{ab}(\mu)$ on the states
$|\l{1}\ldots\l{n}\rangle$ in the limit $\mu\rightarrow\infty$
$$\putequation{lim2}$$
This implies that
$$\putequation{lim3}$$
We now define new operators $J^{(1)}_a$ according to
$$\putequation{j1}$$
{}From \pl{lim3} it is clear that the lowest weight property of the
Bethe Ansatz states with respect to $\tqd_{\pm 1}$ is equivalent to
the conditions
$$\putequation{j1f}$$
Recalling that
$$ F = \prod_{j=1}^{n_1}C^{(1)}_{b_j}(\1l{j})\vacn G^{b_{n_1}\ldots
b_1}$$ we realize that we now face a problem very similar to the one
we just solved for the case of $\eta$ and $Q_{\pm 1}$.
The graded commutator of $J^{(1)}_a$ with $C^{(1)}_b(\1l{})$
can be computed by using \pl{intnest2} and \pl{j1}
$$\putequation{jc1}$$
The action of $J^{(1)}_a$ on the states
$|\1l{1}\ldots\1l{n_1}|G\rangle$ now takes a form very similar to
\pl{lw1}
$$\putequation{nlw1}$$
The computation of the quantities $\Omega^{(1)}_k$ follows the same
strategy employed above for the determination of $\Omega_j$, and after
some computations we find that all $\Omega^{(1)}_k$ vanish, because
the unwanted terms {\sl for the nesting} cancel, {\sl i.e.,}
$$({\tilde\Lambda^{(1)}}_kG)^{b_1\ldots b_{k-1}ab_{k+1}\ldots
b_n}+(\Lambda_k^{(1)}G)^{b_1\ldots b_{k-1}ab_{k+1}\ldots b_n}
=0\qquad .$$
This completes the proof of the lowest weight property for $\tqd_{\pm
1}$.

\subsection{\sl Lowest weight property for $S$}

Noting that
$$\putequation{limS}$$
we find
$$\putequation{limS2}$$
Therefore the condition that $S$ annihilates all Bethe Ansatz states
with finite spectral parameters is equivalent to
$$\putequation{limS3}$$
{}From the intertwining relation \pl{intnest4} we deduce
$$\putequation{limS4}$$
which enables us to show by direct computation that
$$J^{(2)}|\2l{1}\ldots\2l{n_2}\rangle = 0$$ due to the cancellation of
the unwanted terms {\sl for the second nesting}
$$({\tilde\Lambda^{(2)}}_k)+(\Lambda_k^{(2)})=0\qquad .$$
This completes our proof of \pl{lws}.

{\bfs\section{Higher Conservation Laws}}

In this section we derive explicit expressions for the conservation
laws $H_{(3)}$ and $H_{(4)}$ (which involve interactions between $3$
and $4$ neighbouring sites respectively). We use a generalisation of
Tetel'man's method\upref Tetelman, Sklyanin/ to the
supersymmetric case\upref ek/.

Let us define the ``boost''-operator
$$\putequation{boost}$$
where $H_{(2)}^{n,n+1}$ is the density of the hamiltonian given by the
right hand side of (\putlab {trid1}).
This operator obviously violates periodicity on the finite chain, but
if used in commutators (which ``differentiate'' the linear
$n$-dependence) it yields formally periodic expressions.

The integrals of motion can be successively obtained by commutation
with the boost-operator\upref ek/
$$\putequation{hcl2}$$
where
$$\putequation{boost2}$$

Using (\putlab{hcl2}) we obtain explicit expressions for higher
conservation laws as follows~:
We have shown in [\putref{eks}] that the hamiltonian $H^0$ can be written
in terms of $u(2|2)$ generators $\{J_{j,\ga}|\ga = 1\ldots 16\} = \{
S_j,S^\dagger_j,S_j^z,\eta_j,\eta^\dagger_j,\eta_j^z,I_j,X_j,Q_{j,\gs},
Q^\dagger_{j,\gs},{\tilde Q}_{j,\gs},{\tilde
Q}^\dagger_{j,\gs}|\gs=\pm 1\}$ as
$$\putequation{Hagain}$$
The structure functions of $u(2|2)$ are defined by
$$\putequation{f}$$
$H_{(3)}$ can be obtained by commutation with the boost operator
$\tilde B$
$$\putequation{H3}$$
The next highest conservation law can be computed along similar lines
and we find
$$\putequation{H4}$$
where ${P}^{k-1,k+1}$ is a graded permuation operator between the
sites $k-1$ and $k+1$ with definition
$$\putequation{P}$$
\vskip 1cm
\sectionnumstyle{Alphabetic}
\sectionnum=0
\equationnum=0

\begin{appendices}

{\bfs\section{Appendix : Computation of the ``unwanted terms'' for
the FFBB grading}}
In this section we compute the so-called ``unwanted terms'' in the
expressions (\putlab{Dstate}) and (\putlab{Astate}).
These unwanted terms are characterized by containing a creation
operator $C_a$ with spectral parameter (SP) $\mu$ in place of a
creation operator with SP $\l{k}$ (or $\1l{k}$, $\2l{j}$ for the
nestings).
The condition of cancellation of the unwanted terms leads to the
Bethe equations.
In order to obtain the expression for ${\tilde \Lambda}_k$, we first
move the creation operator with SP $\l{k}$ to the first place in
(\putlab{state1}), using
(\putlab{AC})
$$\putequation{front}$$
To get an unwanted term, we now have to use the second term in
(\putlab{AC}) to move $D$ past $C_{b_k}(\l{k})$, and then always the
first term in (\putlab{AC}) to move $D$ (which now carries SP $\l{k}$)
to the very right, until it hits the vacuum, on which it acts according
to (\putlab{Tvac}).
This way we obtain
$$\putequation{laktilde}$$
The computation of $\Lambda_k$ is more complicated. We first derive an
expression for the contribution of $A_{11}(\mu )$, which we denote by
$\Lambda_{k,1}$. Proceeding along the same lines as in the computation
of ${\tilde \Lambda}_k$, we find
$$\putequation{lak1}$$
The $\gd_{d_{n-1},1}$ stems from the action of $A_{1,d_{n-1}}(\l{k})$
(which is what we get after moving $A$ past all the $C$'s) on the vacuum.
We also had to include a $\gd_{b_{k},1}$ due to the fact that in
(\putlab{Astate}) we denoted by $b_k$ the index of the $C$ with SP
$\mu$.
The contributions of $A_{22}(\mu )$  and $A_{33}(\mu )$ can be
obtained in a similar way, and $\Lambda_k =
\Lambda_{k,1}+\Lambda_{k,2}+\Lambda_{k,3}$ is found to be
$$\putequation{lak2}$$
This expression can be simplified by carrying out the
contractions over the summation indices $c_1,\ldots ,c_k$.
Noting that
$$\putequation{invr}$$
we are able to perform all $c_i$-summations with the result
$$\putequation{Sr}$$
Now we transform the remaining $r$-matrices into $L$-operators, using
the identity
$$\putequation{raise}$$
Thus we obtain our final form for the unwanted terms due to
$-A_{11}(\mu )-A_{22}(\mu )+A_{33}(\mu )$
$$\putequation{lak3}$$
We now insert (\putlab{lak3}) and (\putlab{laktilde}) into the
condition (\putlab{unw}) for the cancellation of the unwanted terms,
and multiply the resulting equation by the inverse of $S(\l{k})^{b_1\ldots
b_k}_{a_1\ldots a_k}$, which satisfies $\left(S^{-1}(\l{k})\right)^{p_1\ldots
p_k}_{b_1\ldots b_k}\ S(\l{k})^{b_1\ldots b_k}_{a_1\ldots a_k} =
\prod_{i=1}^k \gd_{a_i,p_i}$, and which is computed {\sl via}
(\putlab{invr}).
After carrying out these manipulations we arrive at
(\putlab{PBC}).\par
The unwanted terms for the nestings can be computed along similar
lines, we refer to [\putref{ek}] for more detailed derivations and
just give the results. For the first nesting we find
$$\putequation{unwn1a}$$
and
$$\putequation{unwn1b}$$
where
$$\putequation{sone}$$

The unwanted terms for the second nesting are
$$\putequation{unwn2}$$
In all the above computations we have used that no two spectral
parameters coincide, even if they belong to different nestings ({\sl
e.g.} $\l{k}\neq \1l{j}\ \forall j,k$). For the Bethe Ansatz without
nesting the wave function vanishes if two spectral parameters
coincide. This is sometimes called ``Pauli principle in rapidity
space''. The implications of coinciding spectral parameters in the
nested Bethe Ansatz is currently under investigation\upref
ek2/.{\bfs\section{Appendix : Computation of the ``unwanted terms'' for
the BBFF grading}}
In this section we compute the unwanted terms in the
expressions (\putlab{Dstate2}) and (\putlab{Astate2}).

To obtain the expression for ${\tilde \Lambda}_kF$ we first
move the creation operator with SP $\l{k}$ to the first place in
(\putlab{state1}), using
(\putlab{AC2})
$$\putequation{front2}$$
We then use the second term in (\putlab{AC2}) to
move $D$ past $C_{b_k}(\l{k})$, and then always the
first term in (\putlab{AC2}) to move $D$ (which now carries SP $\l{k}$)
to the very right, until it hits the vacuum, on which it acts according
to (\putlab{Tvac2}).
We find
$$\putequation{laktilde2}$$
The computation of the unwanted term $\Lambda_kF$ proceeds along the
same lines as in the FFBB case (Appendix A). After some elemetary
algebra we arrive at
$$\putequation{lak22}$$
The summations over $e_1,\ldots , e_k$ can be carried out by using the
identity
$$\putequation{invr2}$$
Evaluation of the condition of cancellation of the unwanted terms
\pl{cancelbff} then leads to \pl{pbcbff1}.
{\bfs\section{Appendix : Equivalence of the BAE}}
In this appendix we establish the equivalence of the six different
sets of BAE by means of particle-hole transformations in rapidity
space\upref Bares2/\footnote{We thank P.A. Bares for explaining this
method to us.}.
Let us denote the various sets of BAE by their underlying grading,
{\sl i.e.} FBBF, BFBF, BBFF, BFFB, FBFB, and FFBB. We first will
demonstrate the equivalence of FBBF and BBFF.\vskip .2cm
\hfill\break
{\sl (i) $FBBF\leftrightarrow BFBF$\hfill}
\vskip .3cm
We start by expressing the third set of (\putlab{BAEFBBF})
$$\putequation{baefbbf3}$$
as a polynomial equation of degree $N_\down +N_l$
$$\putequation{pol}$$
Among the $N_l+N_\down$ roots of (\putlab{pol})
we consider $N_\down$ roots $w_1,\ldots ,w_{N_\down}$,
which we identify with $\ttwol{1},\ldots ,\ttwol{N_\da}$.
The $N_l$ other roots of (\putlab{pol}) we denote by ${w^\prime}_j$.
Using the residue theorem we can derive the following equality
$$\putequation{res}$$
where $C_j$ is a small contour around $w_j$.
The branch cut of the logarithm extends from $z_n=\1l{l}+{i\over
2}$ to $z_p=\1l{l}-{i\over 2}$. By deforming the contours on the
r.h.s. of (\putlab{res}) we arrive the the following equality
$$\putequation{res2}$$
where the last term on the r.h.s. comes from integration around the
branch cut. The form of the polynomial $p$ now implies that
$$\putequation{cut}$$
Inserting (\putlab{cut}) into (\putlab{res2}) and exponentiating the
result we obtain the identity
$$\putequation{id}$$
Now we recall that $w_j=\ttwol{j}$ and use (\putlab{id}) in the second
set of BAE in (\putlab{BAEFBBF}) with the result
$$\putequation{baebfbf2}$$
This is precisely the second set of BAE for the BFBF grading
(\putlab{BAEBFBF}), if we make the identification ${w^\prime}_k =
\ttwol{k}$. The third set of the BAE (\putlab{BAEBFBF}) is also
fulfilled by the spectral parameters $\ttwol{k}$, because they are roots
of the polynomial equation (\putlab{pol}).
The first set of BAE for the BFBF and the FBBF gradings are identical
anyway, so that we thus we have established the equivalence of the BAE
(\putlab{BAEFBBF}) and (\putlab{BAEBFBF}).\vskip .2cm
\hfill\break
{\sl (ii) $BFBF\leftrightarrow BFFB$\hfill}
\vskip .3cm
We start with the first set of the BAE \pl{BAEBFBF} for the BFBF
grading
$$\putequation{baebfbf1}$$
and rewrite it as the polynomial equation
$$\putequation{pol2}$$
Partitioning the roots of (\putlab{pol2}) in the two sets
$\{ w_k|k=1\ldots N_\da+N_b\}=\{ \tl{k}|k=1\ldots N_\da+N_b\}$ and
$\{ w^\prime_j|j=1\ldots L-N_h\}$ and proceeding along the same lines
as in step (i), we can derive the following equality
$$\putequation{id2}$$
Recalling that $w_k=\tl{k}$ and inserting (\putlab{id2}) into the
second set of BAE for the BFBF \pl{BAEBFBF} grading and using that
$L=N_b+N_e$, we obtain
$$\putequation{baebffb}$$
Identifying $w^\prime_j$ with $\tl{j}$ this implies the equivalence of
the BAE (\putlab{BAEBFBF}) and (\putlab{BAEBFFB}).\vskip .2cm
\hfill\break
{\sl (iii) $BFFB\leftrightarrow FBFB$}\hfill
\vskip .3cm
The equivalence of \pl{BAEBFFB} and \pl{BAEFBFB} can be shown by
an analogous computation to the one in step (i).\vskip .2cm
\hfill\break
{\sl (iv) $BFBF\leftrightarrow BBFF$}\hfill
\vskip .3cm
We start by converting the second set of the BFBF BAE \pl{BAEBFBF}
into the polynomial equation
$$\putequation{pol3}$$
We partition the roots of \pl{pol3} into the sets
$\{ w_k|k=1\ldots N_\da+N_l\}=\{ \1l{k}|k=1\ldots N_\da+N_l\}$ and
$\{ w^\prime_j|j=1\ldots N_b\}$. Proceeding along the same lines
as in step (i), we can derive the following equalities
$$\putequation{id3}$$
Inserting \pl{id3} into the BFFB BAE \pl{BAEBFFB} and identifying the
$w^\prime_j$ with $\tonel{j}$ in the BBFF BAE, we see that
\pl{BAEBFFB} and \pl{BAEBBFF} are equivalent.\vskip .2cm
\hfill\break
{\sl (v) $FBFB\leftrightarrow FFBB$}\hfill
\vskip .5cm
This step is analogous to step (iv).\end{appendices}
\newpage
\begin{putreferences}
\centerline{\bfs REFERENCES}
\vskip .5cm
\reference{hirsch}{J.E. Hirsch, \Phy{C158}{1990}{326}.}

\reference{eks}{F.H.L.E\char'31 ler, V.E.Korepin, K.Schoutens,\
\PRL{68}{1992}{2960}.}

\reference{eks2}{F.H.L.E\char'31 ler, V.E.Korepin, K.Schoutens,\
Stony Brook preprint ITP-SB-92-20.}

\reference{yeta}{C.N.Yang,\ \PRL{63}{1989}{2144}.}

\reference{ek}{F.H.L.E\char'31 ler, V.E.Korepin,\ Stony Brook
preprint ITP-SB-92-12, to appear in {\sl Phys.Rev.}{\bfs B}}

\reference{fab}{F.H.L.E\char'31 ler, unpublished .}

\reference{ek2}{F.H.L.E\char'31 ler, V.E.Korepin,\ in preparation.}

\reference{eks3}{F.H.L.E\char'31 ler, V.E.Korepin, K. Schoutens,\ in
preparation.}

\reference{Frahm}{H.Frahm,\ \JPA{25}{1992}{1417}.}

\reference{Sklyanin}{E.K.Sklyanin,\ Univ. of Helsinki preprint
HU-TFT-91-51, Oct.1991}

\reference{Tetelman}{M.G.Tetel'man,\ \JETP{55}{1982}{306}.}

\reference{Bares2}{P.A.Bares, J.M.P.Carmelo, J.Ferrer, P.Horsch,\ MIT
preprint, February 1992.}

\reference{ft}{L.D.Faddeev, L. Takhtajan, Zap.Nauch.Semin LOMI Vol 109
(1981)\ p.134.}

\reference{hubb}{F.H.L.E\char'31 ler, V.E.Korepin, K.Schoutens,\
\NPB{372}{1991}{67},\ \NPB{384}{1992}{431},\ \PRL{67}{1991}{3848}.}

\reference{kdv}{H.J.deVega, M.Karowski,\
\NPB{280}{1987}{225}.}

\reference{fk}{After this work was completed we learned that
A.F\"orster and M.Karowski have proven a highest weight theorem for
the BA states of the supersymmetric $t$-$J$ model with its
$u(1|2)$-symmetry. Their proof is very similar to ours.}

\reference{Cornwell}{J.F. Cornwell, {\it Group Theory in Physics,
Vol III: Supersymmetries and Infinite-Dimensional Algebras}, Academic
Press (1989). The algebras $SU(n|m)$ are discussed on page 270.}

\reference{Kul}{P.P.Kulish,\ \JSM{35}{1985}{2648}.}
\reference{KulRe}{P.P.Kulish, N.Yu. Reshetikhin,\ \JPA{16}{1983}{L591}.}
\reference{KulRe2}{P.P.Kulish, N.Yu. Reshetikhin,\ \JETP{53}{1981}{108}.}
\reference{SklKul}{P.P.Kulish, E.K. Sklyanin,\ \JSM{19}{1982}{1596}.}
\reference{Taktajan}{L.Takhtajan,\ \JSM{23}{1983}{2470}.}
\reference{Vlad}{V.E. Korepin, G. Izergin and N.M. Bogoliubov,
  {\it Quantum Inverse Scattering Method, Correlation Functions
  and Algebraic Bethe Ansatz}, Cambridge University Press, 1992 }

\end{putreferences}
\bye